\documentclass[11pt]{article}

\usepackage{jheppub}

\usepackage{amsmath}
\usepackage{amssymb}
\usepackage{graphicx} 
\usepackage{hyperref}

\usepackage{bm}
\usepackage{bbold}
\usepackage{caption}
\usepackage{subcaption}

\usepackage{verbatim}

\newcommand{\be}{\begin{eqnarray}}
\newcommand{\ee}{\end{eqnarray}}

\newcommand{\bn}{\begin{enumerate}}
\newcommand{\en}{\end{enumerate}}

\def\IZ{\mathbb{Z}}

\def\CA{{\cal A}} 

\def\CC{{\cal C}}

\def\CF{{\cal F}}

\def\CI{{\cal I}}

\def\CL{{\cal L}}

\def\CN{{\cal N}}
\def\CO{{\cal O}}

\def\CS{{\cal S}}
\def\CT{{\cal T}}

\def\a{\alpha}

\def\e{\epsilon}

\def\s{\sigma}

\def\G{\Gamma}

\def\half{\frac{1}{2}}

\def\goto{\rightarrow}

\def\p{\partial}

\def\Tr{{\rm Tr}}
\def\tr{{\rm Tr}}

\def\PE{{\rm PE}}
\def\vec#1{\bm{#1}}

\title{New Dualities and Misleading Anomaly Matchings from Outer-automorphism Twists}

\author{Sridip Pal}
\author{and Jaewon Song}

\affiliation{Department of Physics, 
University of California, San Diego\\
La Jolla, CA 92093, USA}

\emailAdd{srpal@ucsd.edu}
\emailAdd{jsong@physics.ucsd.edu}

\abstract{We study four-dimensional $\CN=1, 2$ superconformal theory in class $\CS$ obtained by compactifying the 6d $\CN=(2, 0)$ theory on a Riemann surface $\CC$ with outer-automorphism twist lines. From the pair-of-pants decompositions $\CC$, we find various dual descriptions for the same theory having distinct gauge groups. We show that the various configurations of the twist line give rise to dual descriptions for the identical theory. We compute the 't Hooft anomaly coefficients and the superconformal indices to test dualities. Surprisingly, we find that the class $\CS$ theories with twist lines wrapping 1-cycles of $\CC$ have the identical 't Hooft anomalies as the ones without the twist line, whereas the superconformal indices differ. This provides a set of examples where the anomaly matching is insufficient to test dualities. 
}
	
\makeindex

\begin{document}
\maketitle

\section{Introduction}

One of the most important discoveries in supersymmetric field theory is that there can be two different theories flowing to the same infrared (IR) fixed point \cite{Seiberg:1994pq}. This IR duality has been extensively studied for the last two decades. However, the test of the duality is rather difficult in general since at least one of the dual descriptions is strongly-coupled in the IR. The usual test involves quantities that can be computed reliably due to the holomorphic nature: such as matching of chiral operators, moduli space of vacua and the 't Hooft anomalies. One of the most useful quantities to compute is the 't Hooft anomaly coefficients as they are invariant under the renormalization group (RG) flow, and hence can be computed in the weakly coupled regime. Thus, two very different gauge theories having the same anomaly coefficients is a quantitative check of the duality. 

The two different theories having the same 't Hooft anomalies, in spite of being a strong check, does not necessarily imply that two theories are dual to each other. After all, they are small set of numbers one can compute from the theory, so that the matching can be a coincidence. Indeed such example was given in \cite{Brodie:1998vv}. But such examples are rather difficult to find because it involves understanding of strong-coupling physics at the fixed point. 

Recent developments in the supersymmetric localization provide us with new tools for testing dualities. This gives new set of physical observables, e.g. supersymmetric partition functions, that can reliably be computed. These supersymmetric partition functions allow us to extract certain operator spectrum of the fixed point, and also provide more refined checks for the dualities. See the recent review \cite{Pestun:2016zxk} and the references therein for more details. So it is a natural question to ask whether there can be a pair of theories with the same 't Hooft anomalies but different supersymmetric partition functions. 

In this paper, we present a set of examples that have the same anomaly coefficients but different supersymmetric partition functions (namely the superconformal index). This set of examples arises from the so-called class $\CS$ theories \cite{Gaiotto:2009we, Gaiotto:2009hg}, coming from the 6d $\CN=(2, 0)$ theory compactified on a Riemann surface $\CC_{g, n}$ of genus $g$ with $n$ punctures. The Riemann surface is called the UV curve, and one can get $\CN=1$ or $\CN=2$ theory in four-dimension depending on the choice of partial topological twist along the UV curve. One of the powerful features of the class $\CS$ theory is that various properties become manifest through the underlying geometry of $\CC_{g, n}$. Especially, different pair-of-pants decompositions give various dual descriptions. In this way, new $\CN=2$ S-dualities \cite{Gaiotto:2009we, Tachikawa:2009rb,Tachikawa:2010vg, Chacaltana:2010ks,Chacaltana:2011ze, Nishinaka:2012vi, Chacaltana:2012ch,Chacaltana:2013oka,Chacaltana:2014jba,Chacaltana:2014nya,Chacaltana:2015bna,Chacaltana:2016shw} and $\CN=1$ dualities have been found \cite{Benini:2009mz,Bah:2012dg, Gadde:2013fma,Xie:2013gma, Bah:2013aha, Agarwal:2013uga, Agarwal:2014rua,Agarwal:2015vla}. 

The setup we consider in this paper is the class $\CS$ theories of type $\G=ADE$ obtained by 6d $\CN=(2, 0)$ theory of type $\G$ wrapped on a UV curve $\CC_{g, n}$ of genus $g$ and $n$ punctures. Let us denote the 4d theory obtained in this way as $\CT[\CC_{g, n}]$.\footnote{Here we suppress the reference to the type $\G$ of the 6d $\CN=(2, 0)$ theory and the type of each punctures.} When the group $\G$ admits an outer-automorphism action, one can introduce an outer-automorphism twist line, or twisted punctures \cite{Tachikawa:2010vg, Chacaltana:2012ch,Chacaltana:2013oka,Agarwal:2013uga, Chacaltana:2015bna,Chacaltana:2016shw}.\footnote{We only discuss the $\IZ_2$-twist in this paper. When $\G=D_4$, one can also twist with the $\IZ_3$ action. Most of the story regarding the anomalies vs indices holds for this case as well.} This twisting can be understood as introducing an orientifold plane in the type IIA intersecting brane description, which changes the gauge group and the matter content of the 4d theory appropriately. The twist lines can emanate from the twisted puncture and end on another twisted puncture. The twist line can also form a loop wrapping various 1-cycles of $\CC_{g, n}$ \cite{Nishinaka:2012vi}. Let us call the latter ones as the twist loops. One can have various  configurations of the twist loop for a given UV curve. Surprisingly, it turns out that various configurations for the twist lines can be deformed to each other, so that we find:
\begin{quote}
there is only one \emph{physically inequivalent} configuration for the $\IZ_2$ twist loop. 
\end{quote}
Different topological configurations give rise to dual descriptions of the same theory upon pair-of-pants decomposition.\footnote{The twist lines define a $\IZ_2$-bundle over the UV curve $\CC$. The topological class is captured by its Stiefel-Whitney class, which takes value in $H^2 (\CC, \IZ_2)=\IZ_2$. This is consistent with what we find. We would like to thank Yuji Tachikawa for pointing this out to us.}

Also, when there is a twisted-puncture on the UV curve, we find that adding a twist loop to the curve does not change the theory but gives a dual description:
\begin{align}\label{dual}
 \CT[\CC_{g, n}] \simeq \CT[\tilde{\CC}_{g, n}] \qquad \textrm{(if any of the punctures are twisted)} \ , 
\end{align}
where $\tilde{\CC}_{g, n}$ denotes the same UV curve as $\CC_{g, n}$ except for a twist line wrapping around 1-cycle. It is rather surprising on first sight, but it turns out the two configurations can be obtained by deforming the geometry in a smooth way. 

We compute the anomaly coefficients and the superconformal indices to test the dualities. Since the theories that we discuss in this paper are mostly `non-Lagrangian', we rely on the 4d/2d correspondence of indices for the class $\CS$ theories \cite{Gadde:2009kb,Gadde:2011ik,Gadde:2011uv, Gaiotto:2012xa, Mekareeya:2012tn, Lemos:2012ph, Beem:2012yn}. We find that the anomalies for the class $\CS$ theories does not depend on the existence of the twist loop, but the theories belong to different universality classes, provided none of the punctures are twisted:
\begin{align}
 \CA \left( \CT[\CC_{g, n}] \right) = \CA ( \CT[\tilde{\CC}_{g, n}] ) \quad \textrm{whereas} \quad
 \CT[\CC_{g, n}]  \neq \CT[\tilde{\CC}_{g, n}] 
\end{align}
where $\CA(\cdot)$ denotes the anomaly coefficients of the theory. Even though the anomalies match, these two theories differ, which can be checked from distinct superconformal indices. If we have at least one twisted puncture, we obtain the new dualities as in \eqref{dual}.

The organization of this paper is as follows. In section \ref{sec:duality}, we describe the new dualities we obtain by considering class $\CS$ theories with outer-automorphism twist loops on the UV curve. In section \ref{sec:anomaly}, we compute the 't Hooft anomaly coefficients and the superconformal indices to test the dualities. We find that the anomaly coefficients does not get altered by the presence of loop of the twist line, but the index gets modified. Then we conclude with some remarks. 


\section{New dualities from the outer-automorphism twist} \label{sec:duality}
\subsection{$\CN=2$ dualities}
\paragraph{$\CN=2$ theories of class $\CS$}
Let us review aspects of the class $\CS$ theories \cite{Gaiotto:2009we, Gaiotto:2009hg} that we need in our discussion. The class $\CS$ theories arise from wrapping 6d $\CN=(2, 0)$ theory of type $\G \in ADE$ on a Riemann surface $\CC$ with or without appropriate punctures. The punctures arise from various codimension 2 defects of the 6d theory. In order to preserve some amount of supersymmetry, one should perform partial topological twist along the direction of $\CC$. Depending on the choice of embedding $SO(2)$ holonomy group into the $SO(5)_R$ of the 6d theory, we get either $\CN=1$ or $\CN=2$ supersymmetry in four-dimensions \cite{Bah:2012dg, Xie:2013gma}.  Let us first focus on $\CN=2$ case for the moment. 

\begin{table}[t]
\centering
\begin{tabular}{c|cccc}
 $\G$ & $A_{2n}$ & $A_{2n-1}$ & $D_{n+1}$ & $E_6$ \\
 \hline
 $G$ & $C_n$ & $B_n$ &  $C_n$  & $F_4$
\end{tabular}
\caption{$\IZ_2$ Outer-automorphisms of the $ADE$ group. Here $G$ is the group obtained from folding the Dynkin diagram of $\G$. }
\end{table}

\begin{figure}[!h]
\centering
\begin{subfigure}[c]{3in}
\centering
\begin{tabular}{cc}
 \includegraphics[height=1.2in]{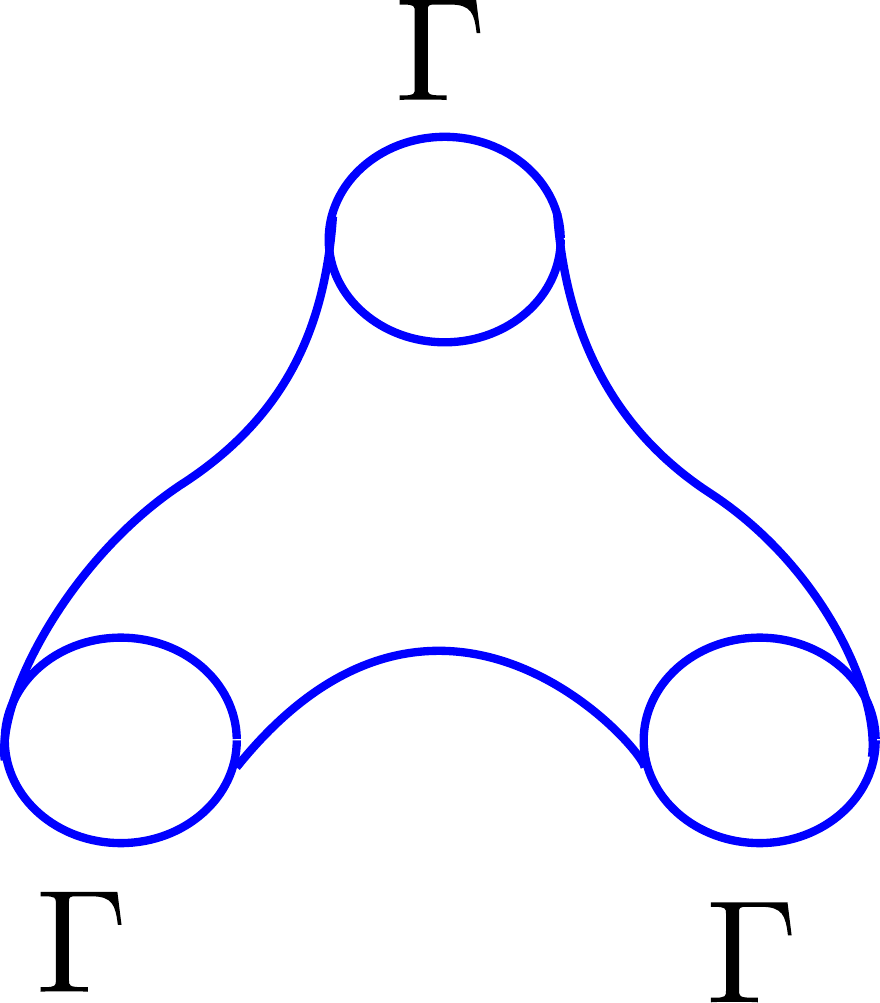} & 
 \includegraphics[height=1.3in]{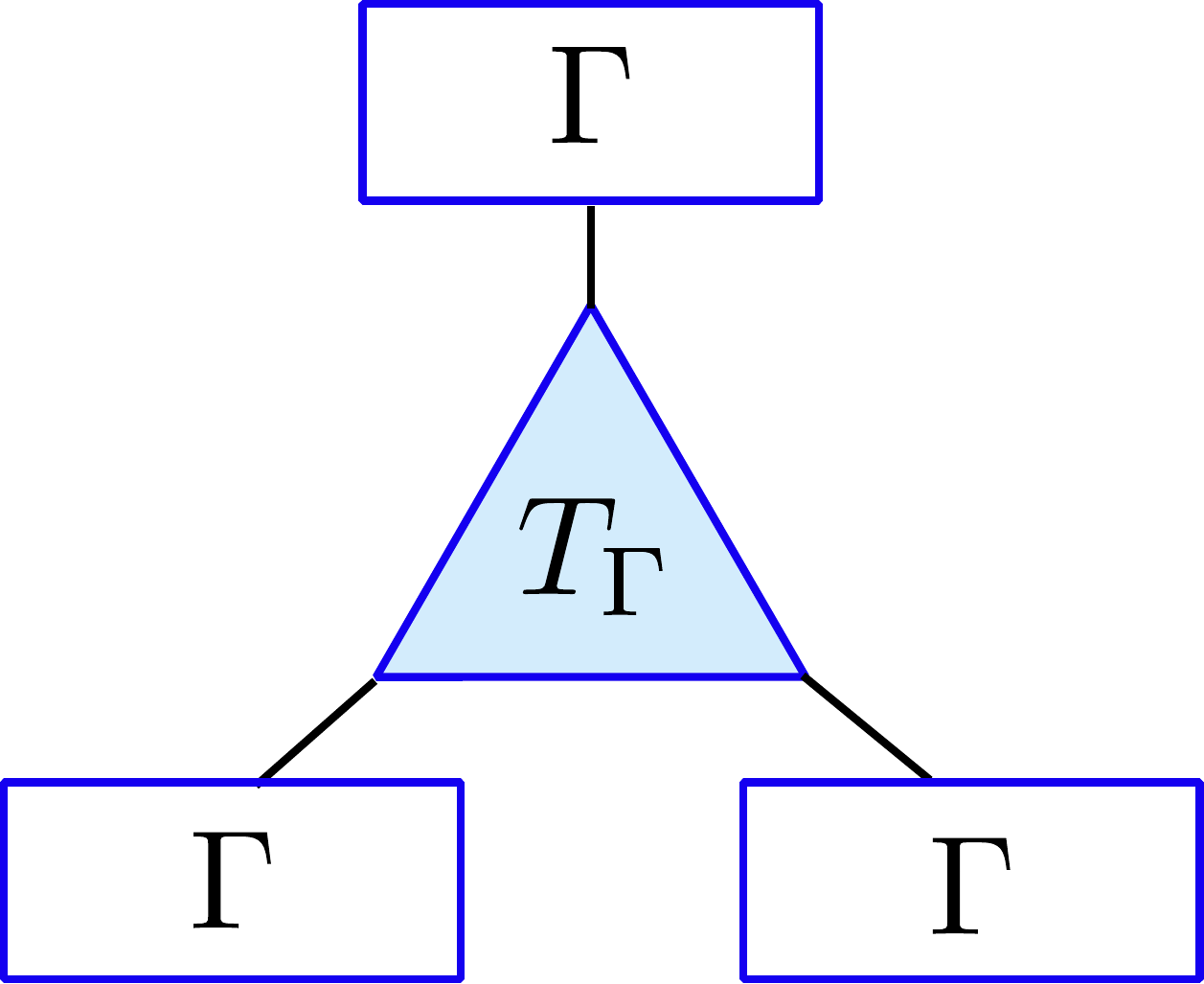}
\end{tabular}
 \caption{A pair-of-pants: $T_\G$}
\end{subfigure}
\begin{subfigure}[c]{3in}
\centering
\qquad
\begin{tabular}{cc}
 \includegraphics[height=1.2in]{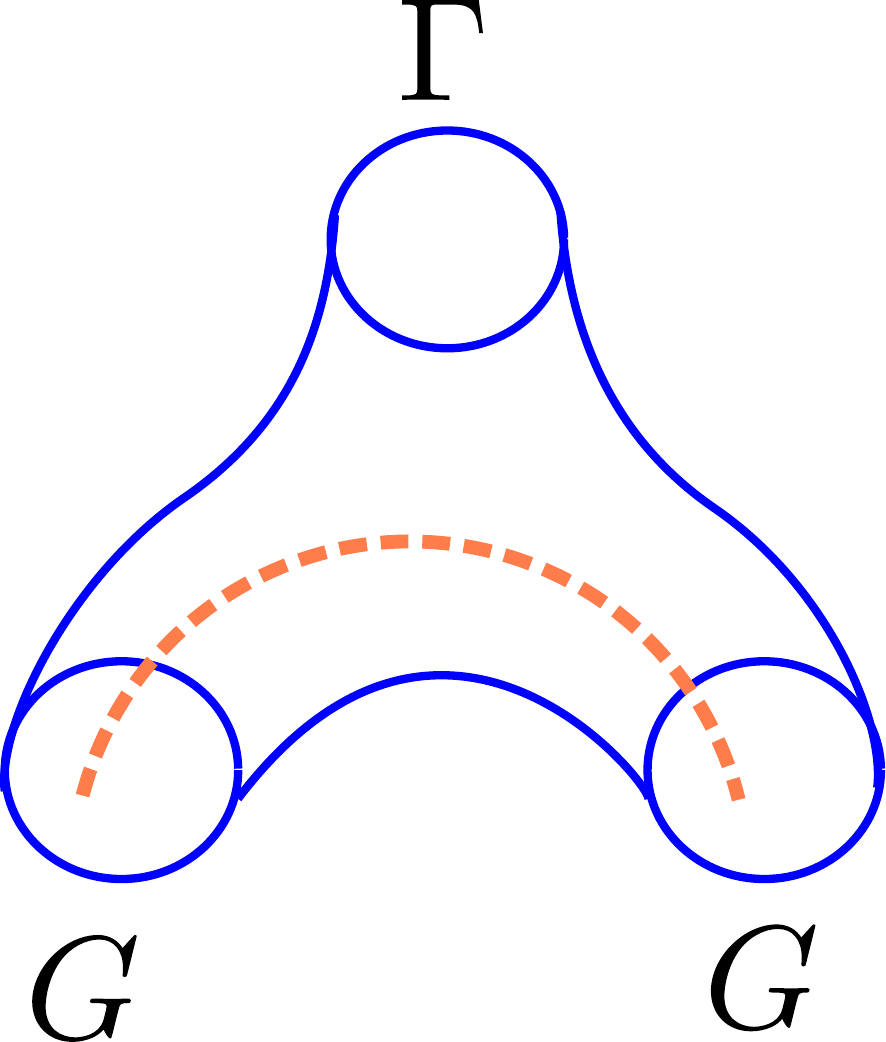} & 
 \includegraphics[height=1.3in]{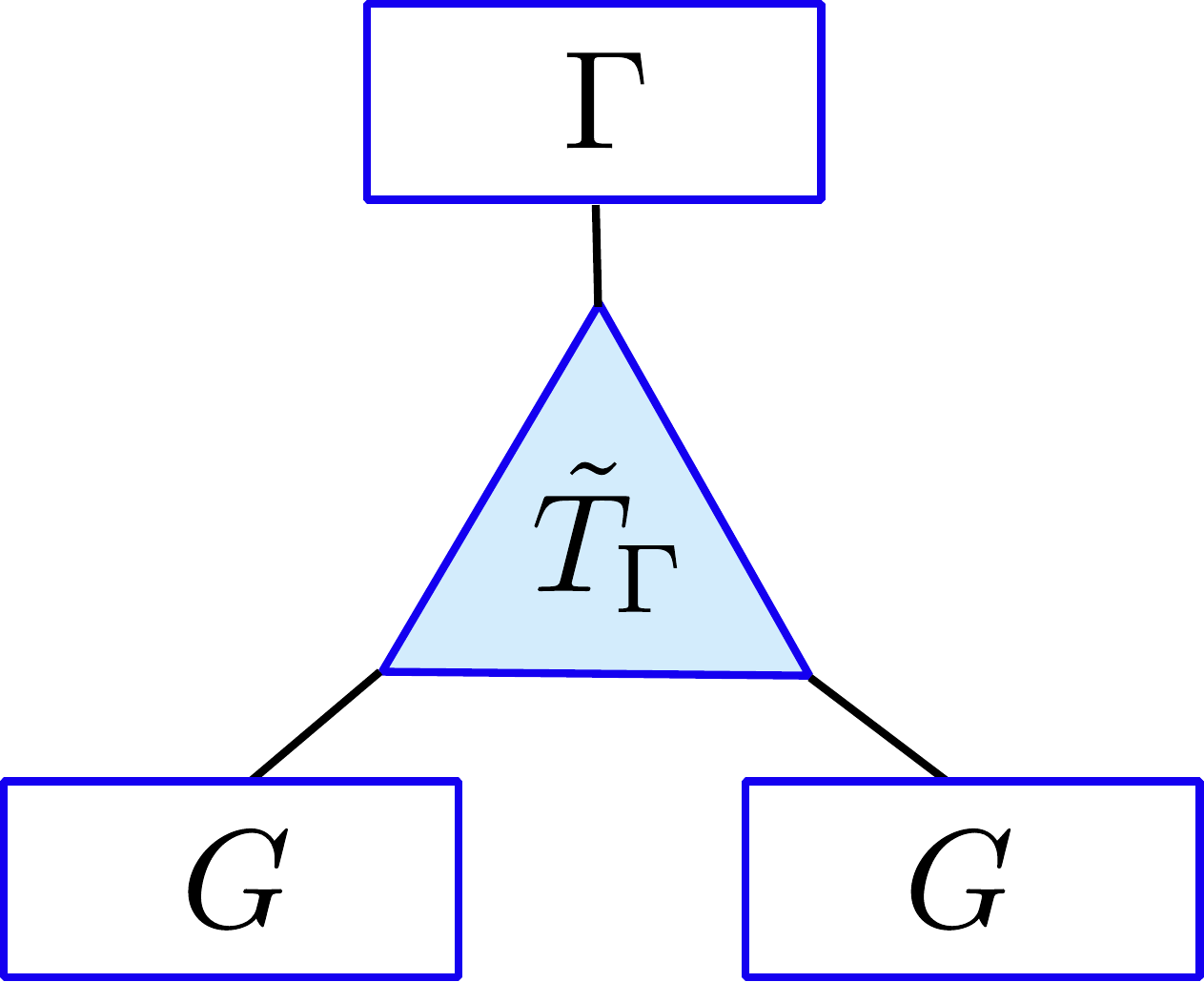}
\end{tabular}
\caption{A pair-of-pants with a twist line: $\tilde{T}_\G$}
\end{subfigure}
 \caption{Two basic building blocks; $T_\G$ and $\tilde{T}_\G$. The twist line (orange) connects between the two twisted punctures with the flavor symmetry $G$.}
 \label{basicbuildingblock}
\end{figure}

The field theory represented by the UV curve is obtained by considering a pair-of-pants decomposition of $\CC$, which is the building block for any (negatively curved) Riemann surface. To each pair-of-pants or the 3-punctured sphere, one associate the $T_\Gamma$ theory, which is an $\CN=2$ superconformal theory with $\G \times \G \times \G$ flavor symmetry. See \cite{Tachikawa:2015bga} and the references therein for more detailed information about this theory. To each puncture\footnote{They are maximal punctures. Non-maximal puncture can be obtained via certain nilpotent Higgsing.} of the 3-punctured sphere, we associate a flavor symmetry $\G$. 
When $\G$ admits an $\IZ_2$ outer-automorphism\footnote{We only consider $\IZ_2$ outer-automorphism. When $\G=D_4$, one can also consider the $\IZ_3$ twist.}, one can also consider a variation of the $T_\Gamma$ theory, which we call $\tilde{T}_\G$ \cite{Tachikawa:2009rb,Tachikawa:2010vg, Chacaltana:2012ch,Chacaltana:2013oka, Agarwal:2013uga, Chacaltana:2015bna,Chacaltana:2016shw}. It is also an $\CN=2$ SCFT, but with the flavor symmetry $\G \times G \times G$, where $G$ is the group associated to the $\G$ by `folding' the associated Dynkin diagram with respect to the outer-automorphism action. In this case, two of the punctures in a pair-of-pants become `twisted', and carry the flavor symmetry $G$. The basic building blocks are drawn in figure \ref{basicbuildingblock}.

\begin{figure}[!h]
\begin{subfigure}[c]{0.5\textwidth}
\begin{center}
 \includegraphics[width=0.7\textwidth]{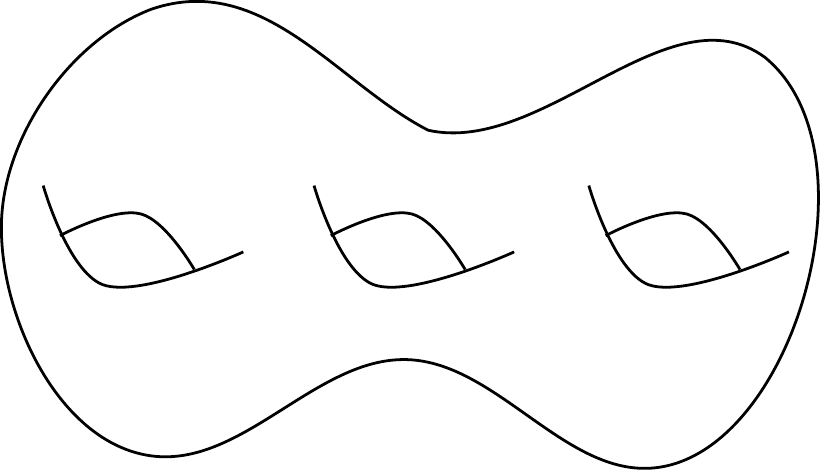}
 \end{center}
 \caption{$\CC_{3,0}$ : UV curve without twist loop}
 \end{subfigure}
 \begin{subfigure}[c]{0.5\textwidth}
 \begin{center}
 \includegraphics[width=0.7\textwidth]{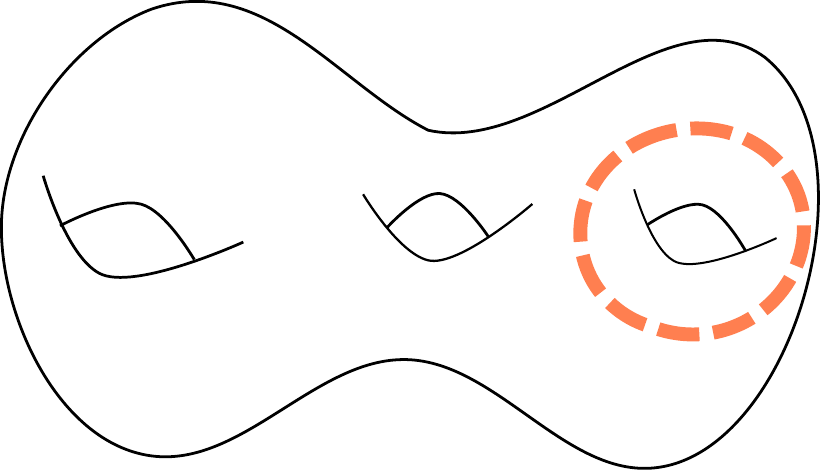}
\end{center}
 \caption{$\tilde{\CC}_{3,0}$ : UV curve with a twist loop}
\end{subfigure}
\caption{The twist line (orange) wraps around one hole. The diagram depicts the case $g=3$.}
\label{fig:genus3}
\end{figure}
\begin{figure}[!h]
\begin{subfigure}[c]{0.5\textwidth}
\centering
 \includegraphics[width=0.9\textwidth]{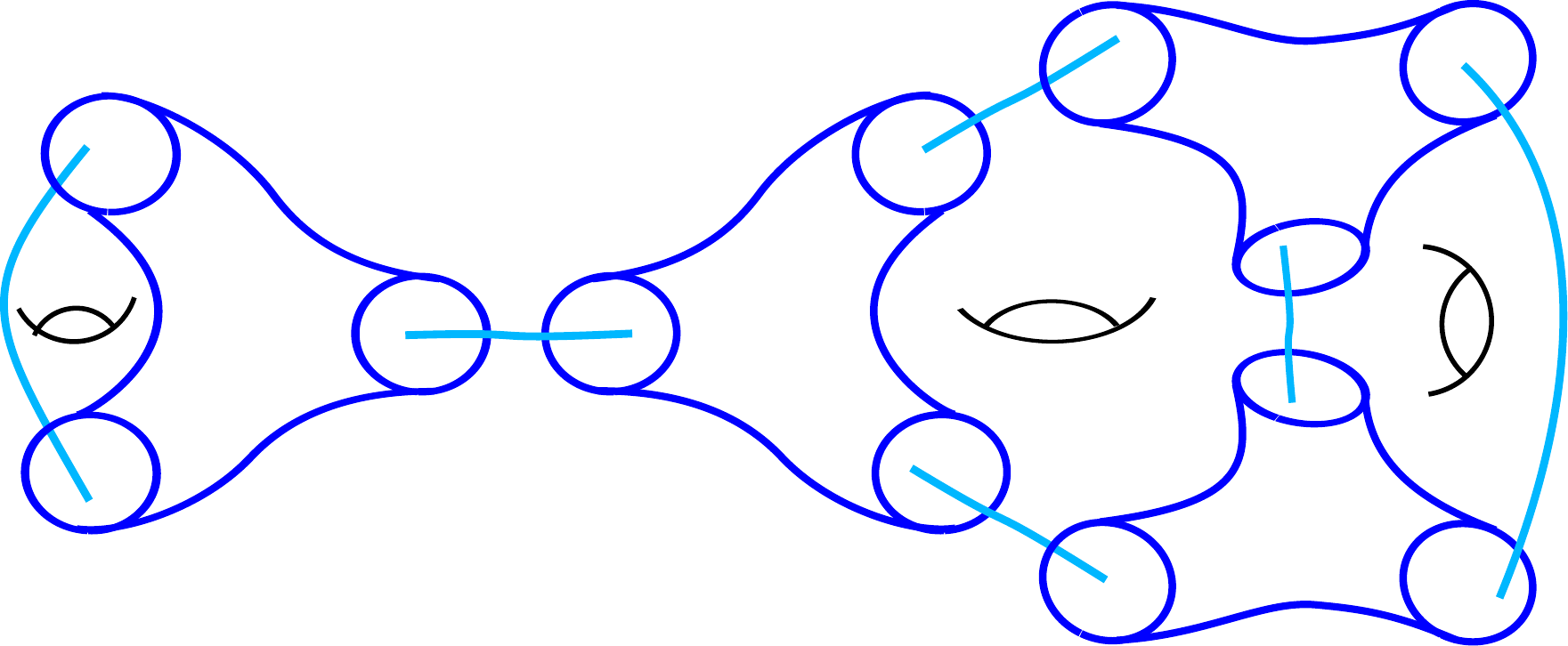}
 \caption{A pair-of-pants decomposition of $\CC_{3,0}$}
 \end{subfigure}
 \begin{subfigure}[c]{0.5\textwidth}
 \centering
 \includegraphics[width=0.9\textwidth]{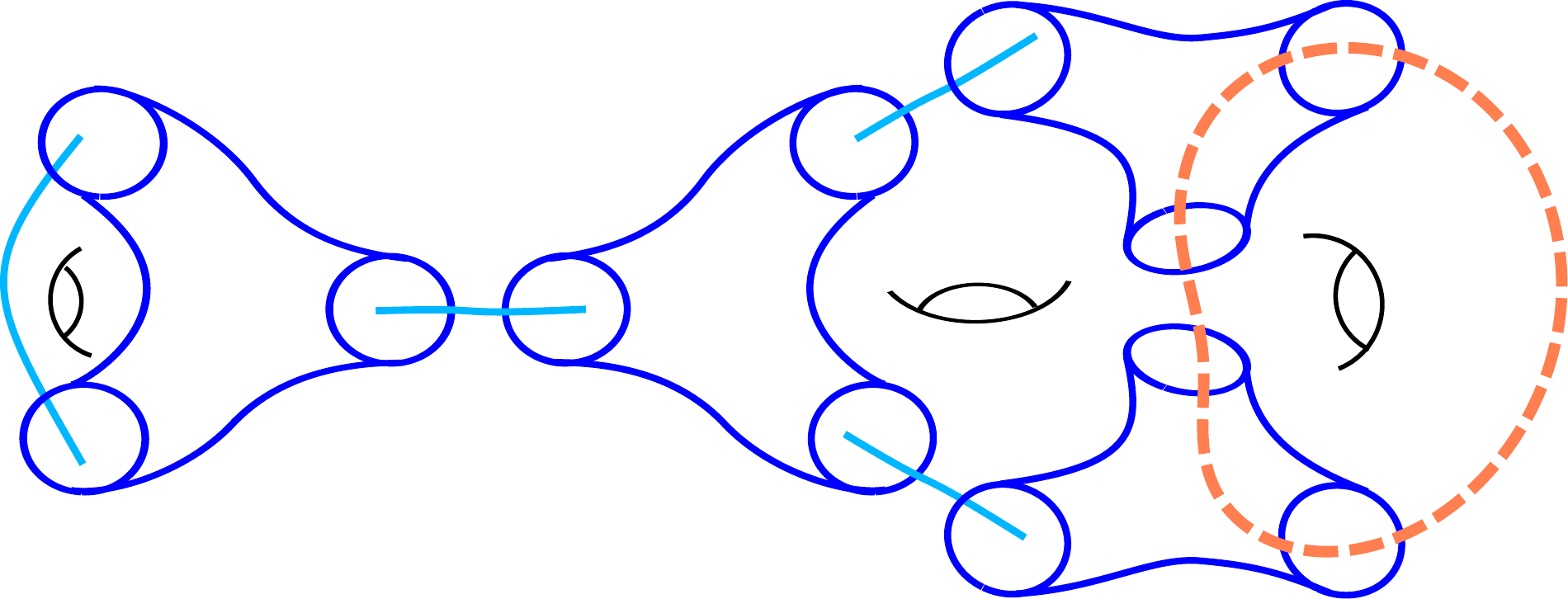} 
 \caption{A pair-of-pants decomposition of $\tilde{\CC}_{3,0}$}
\end{subfigure}
\caption{Pair-of-pants decompositions of $\CC_{3,0}$ and $\tilde{\CC}_{3, 0}$. The blue line represents $\G$ gluing. The orange line represents the $\IZ_2$-twist line, which up on gluing we get the gauge group $G$. }
\label{fig:g3PoP}
\end{figure}

A Riemann surface $\CC$ can be obtained by gluing a number of pair-of-pants (or 3-punctured sphere). The gluing operation is translated to the gauging of the flavor symmetry associated with the punctures. Since a Riemann surface admits various pair-of-pants decompositions, we have one-to-many correspondence. The different choices give the dual descriptions for the same theory. See the figure \ref{fig:genus3} and \ref{fig:g3PoP}. 

\begin{figure}[!h]
\begin{subfigure}[c]{\textwidth}
\begin{center}
\begin{tabular}{cc}
 \includegraphics[height=0.14\textwidth]{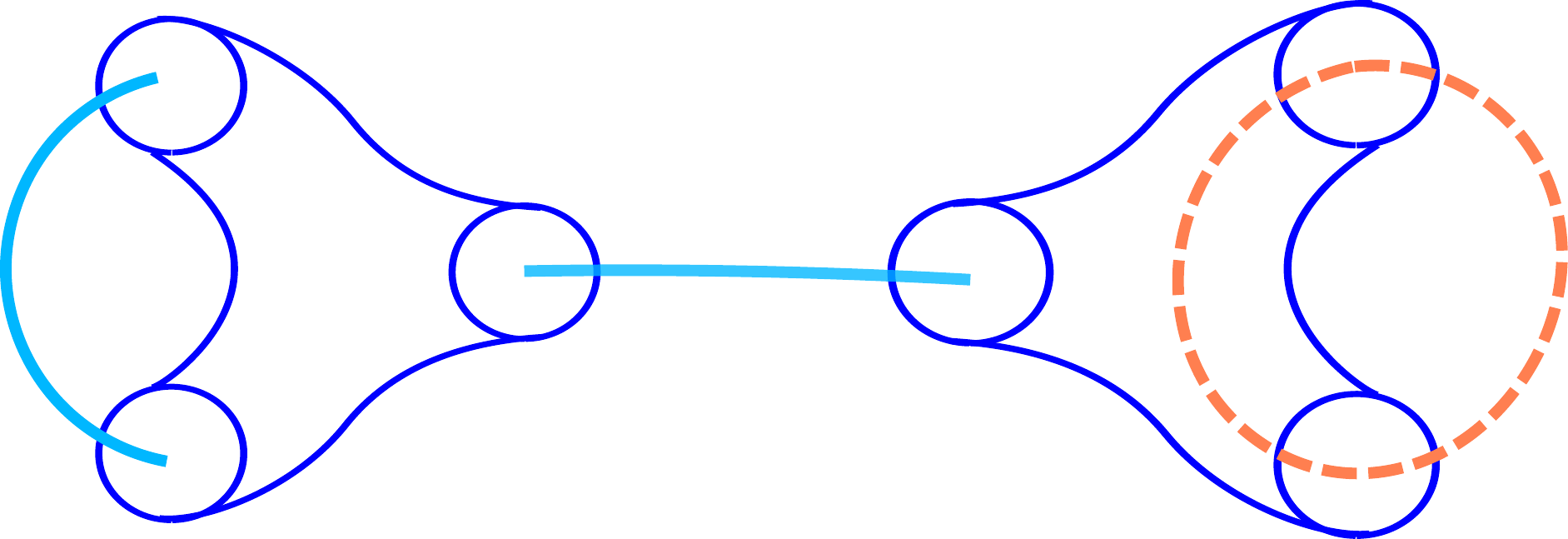} &
 \hspace{0.3in}
 \includegraphics[height=0.14\textwidth]{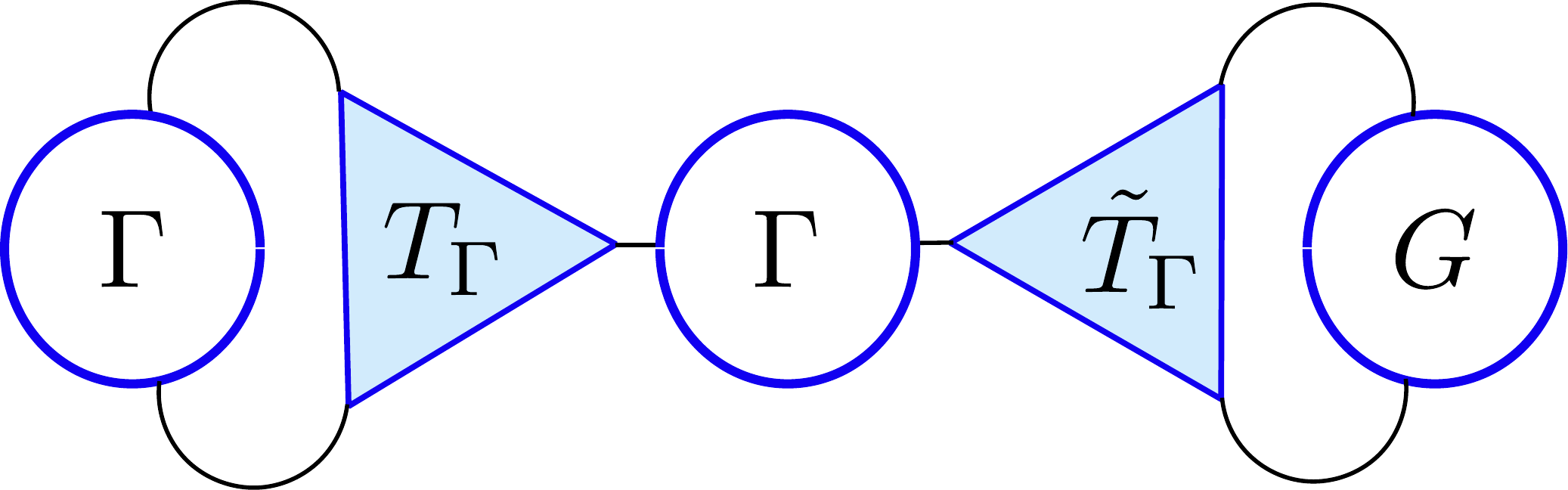}
 \end{tabular}
 \end{center}
 \caption{A pair-of-pants decomposition.}
 \end{subfigure}
 \begin{subfigure}[c]{\textwidth}
 \begin{center}
  \includegraphics[width=0.2\textwidth]{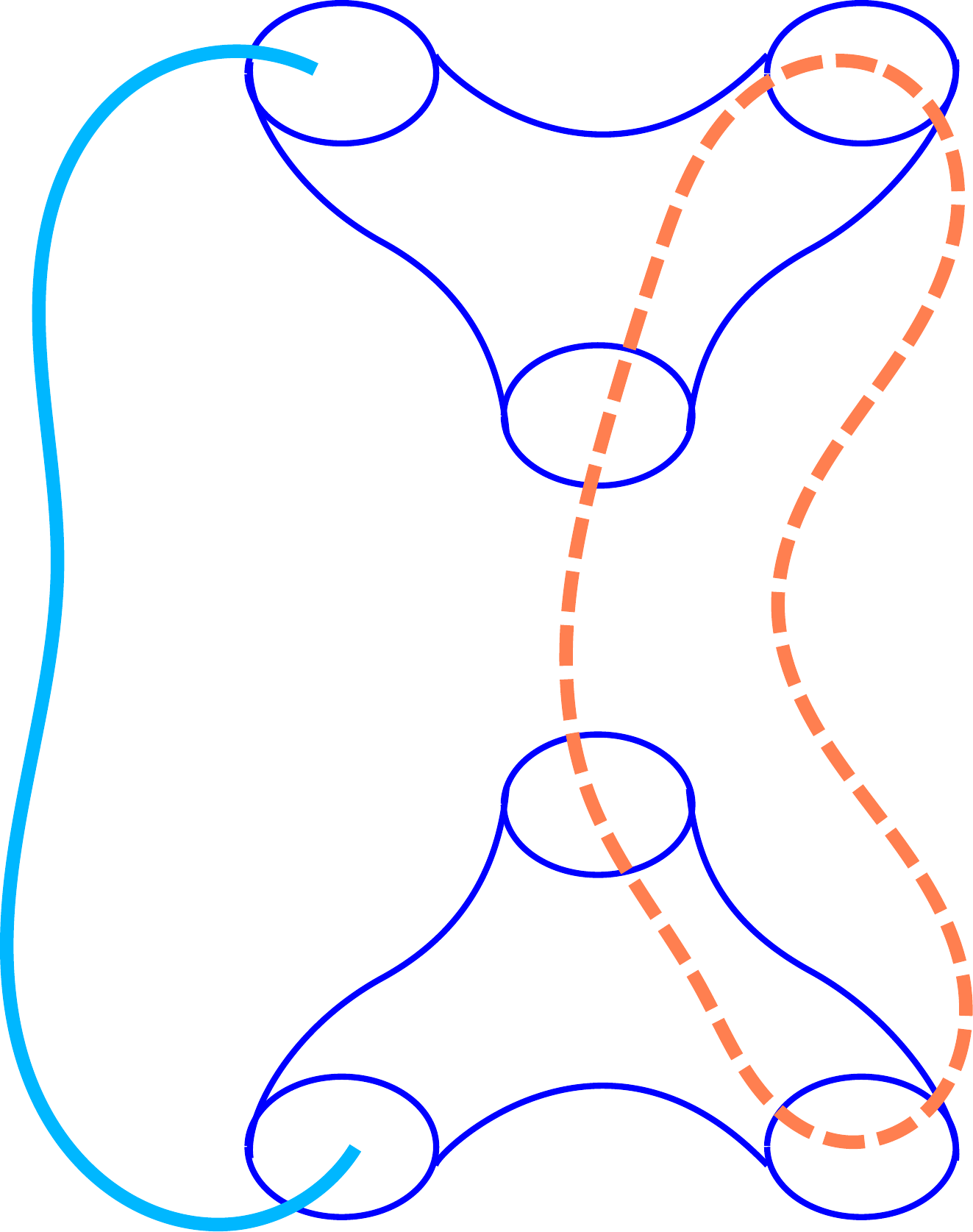}
  \hspace{1in}
  \includegraphics[width=0.25\textwidth]{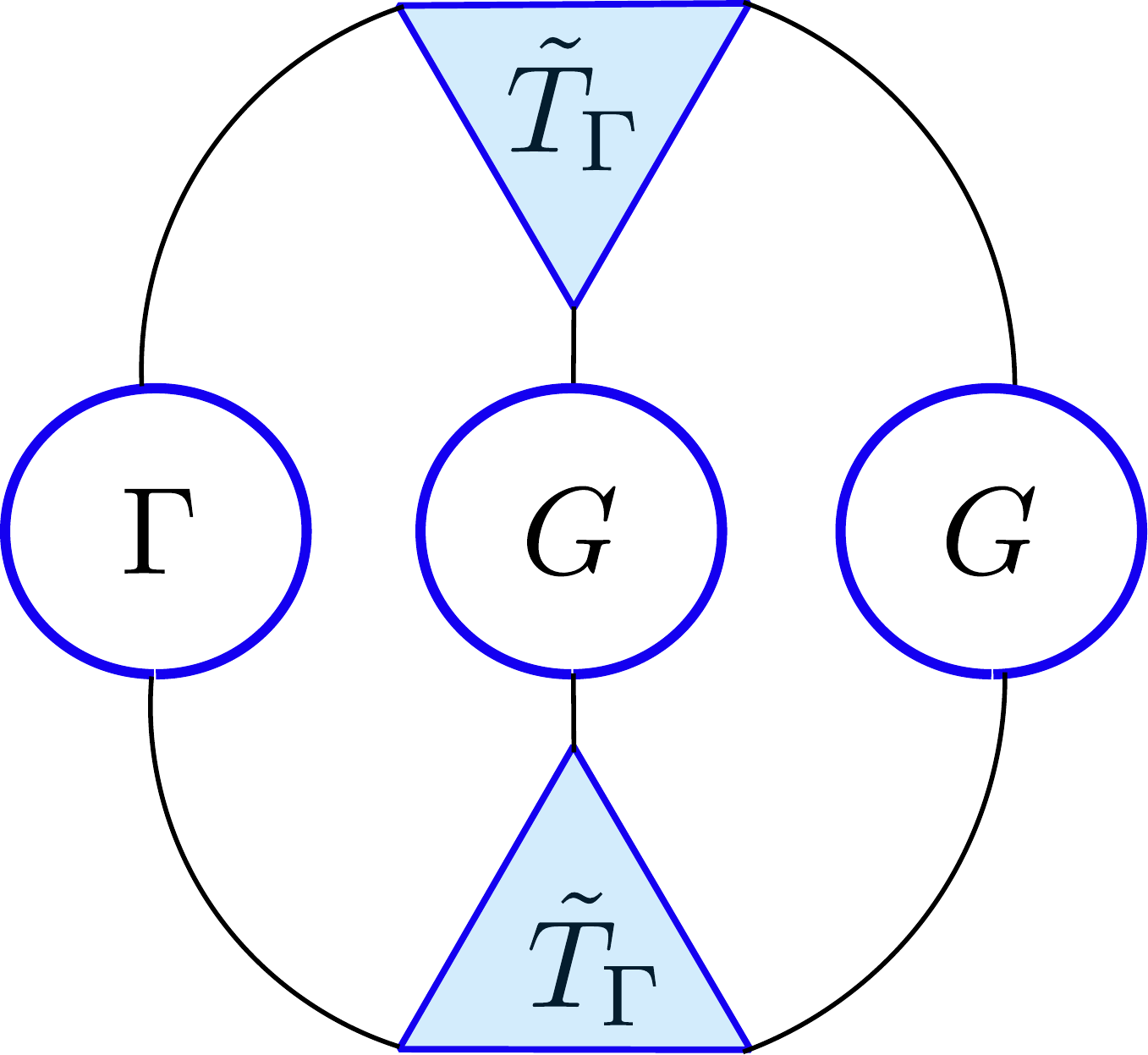}
\end{center}
\caption{Another pair-of-pants decomposition.}
\end{subfigure}
\caption{Two ways of decomposing the UV curve $\tilde{\CC}_{2,0}$ (labelled as $a$ and $b$). Orange line is the twist line. Blue lines represent gluing with $\G$ group. Orange line represent gluing with the $G$ (twisted) gauge group.}
\label{fig:g2PoP}
\end{figure}

\paragraph{Warm up: $g=2$ duality}
Let us first consider the case of the UV curve $\tilde{\CC}_{2,0}$ of genus 2 with a twist line wrapping around a hole. We can decompose the UV curve in terms of the pairs-of-pants in two distinct ways as given in the LHS of the figure \ref{fig:g2PoP}. 

These two different pair-of-pants decompositions will lead to two dual descriptions as given in the RHS of figure \ref{fig:g2PoP}. 
The first one will yield a $T_\G$ block and a $\tilde{T}_\G$ block with $\G \times \G \times G$ gauge symmetry, while the second decomposition will yield two $\tilde{T}_\G$ blocks with $ \G \times G \times G$ gauge symmetry. 
Because they come from the same Riemann surface, the two pair-of-pants decompositions should describe the same physics.
This particular duality can be thought of as a corollary of the basic S-duality for the 4-punctured sphere theory (analogous to the crossing symmetry) formed by gluing $T_\G$ or $\tilde{T}_\G$ \cite{Gaiotto:2009we,Tachikawa:2009rb, Agarwal:2013uga}. We can form the genus 2 surface by gluing the punctures, which is equivalent to gauging the corresponding flavor symmetry. In this way we obtain the two configurations as in the figure \ref{fig:g2PoP}. As we will show in section \ref{sec:N2cc} and \ref{sec:N2idx}, the central charges and the superconformal indices for the two dual frames are identical.

\begin{figure}[!h]
\begin{center}
\begin{subfigure}[b]{2in}
\begin{center}
\includegraphics[width=2in]{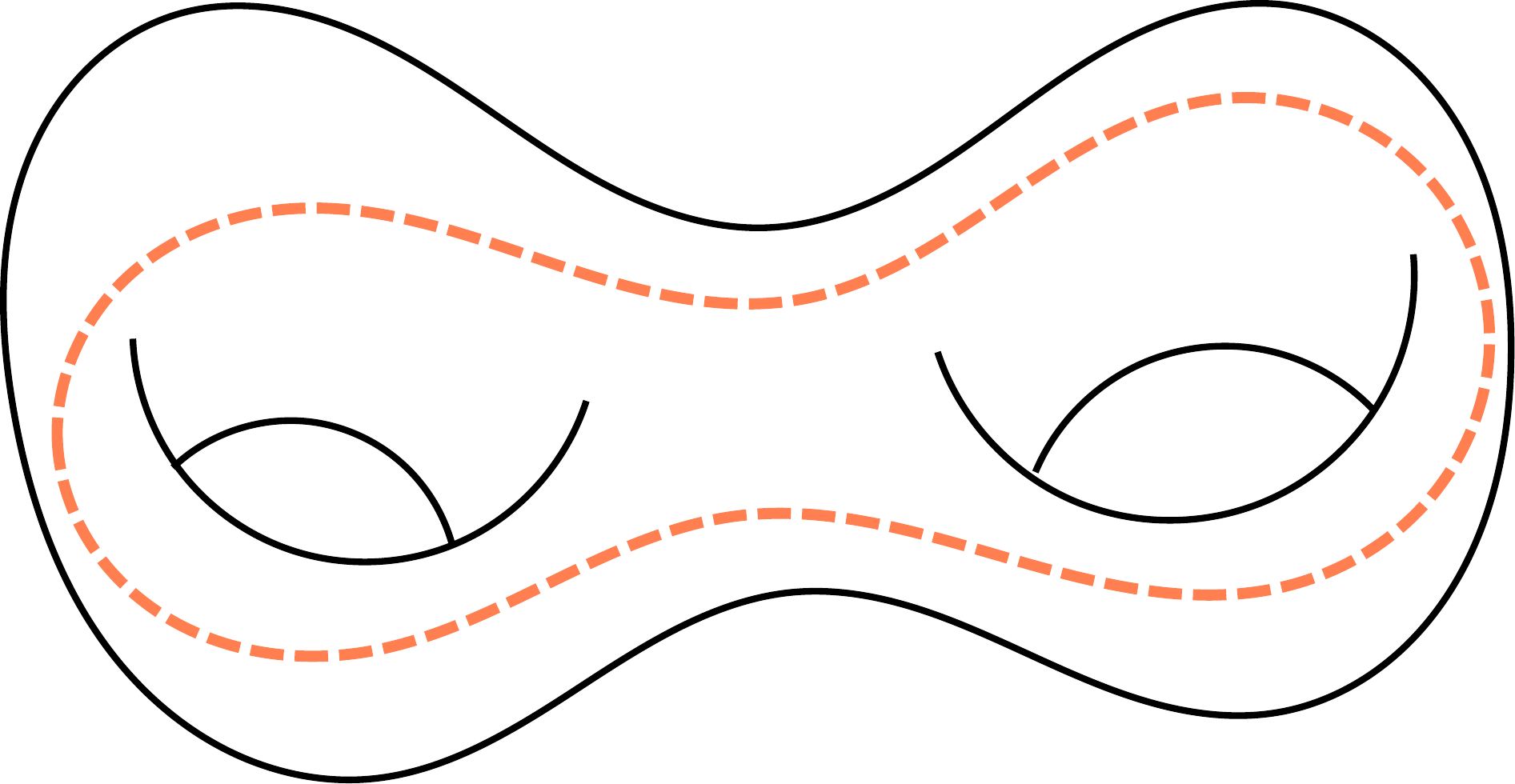}
\vspace{0.05in}
\end{center}
\caption{The UV curve $\tilde{\CC}'_{2,0}$}
\end{subfigure}
\hspace{1in}
\begin{subfigure}[b]{2.2in}
\begin{center}
 \includegraphics[width=1.3in]{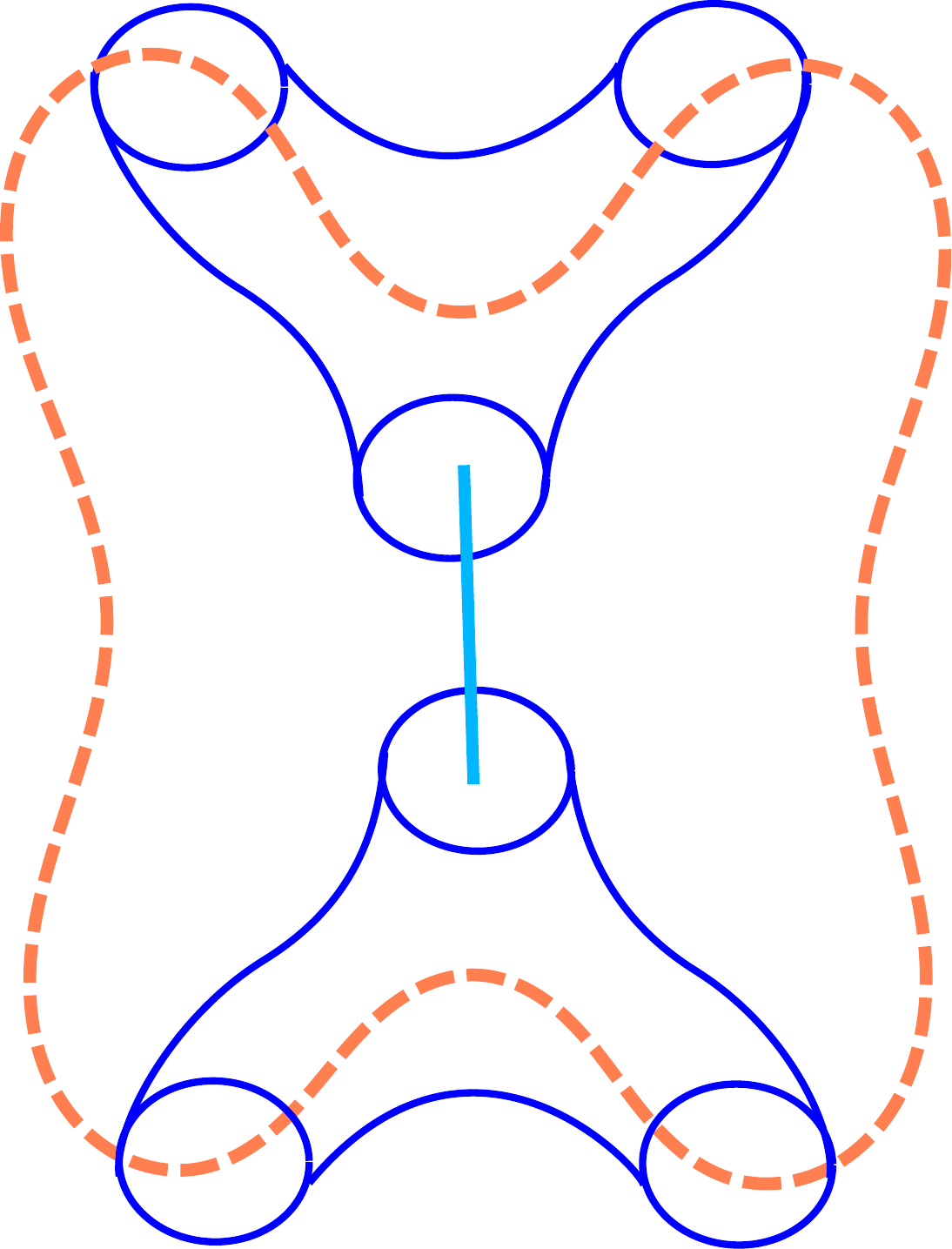}
 \end{center}
 \caption{A pair-of-pants decomposition.}
 \end{subfigure}
 \end{center}
 \caption{The UV curve $\tilde{\CC}'_{2, 0}$. The twist line (orange) wraps around two holes. The pair-of-pants decomposition reveals that we get the same theory as the case where the twist line wraps only one hole.}
 \label{wrap2hole}
\end{figure}
Now, let us consider the situation where the twist line wraps around the two holes at the same time as in figure \ref{wrap2hole}. Let us call this UV curve as $\tilde{\CC}'_{2, 0}$. Note that $\tilde{\CC}_{2, 0}$ and $\tilde{\CC}'_{2, 0}$ are topologically distinct since the twist line cannot be continuously deformed to the other shape. As before, consider a pair-of-pants of decomposition. We get two $\tilde{T}_\G$ blocks with $\G \times G\times G$ symmetry, the same dual frame as we obtained from the second pair-of-pants decomposition of $\tilde{\CC}_{2,0}$ in figure \ref{fig:g2PoP}. They are the same since there is a permutation symmetry among the punctures for the $T_\G$ and $\tilde{T}_\G$ theory. 
\begin{figure}[!h]
\begin{center}
\begin{subfigure}[b]{3in}
\begin{center}
\includegraphics[width=2in]{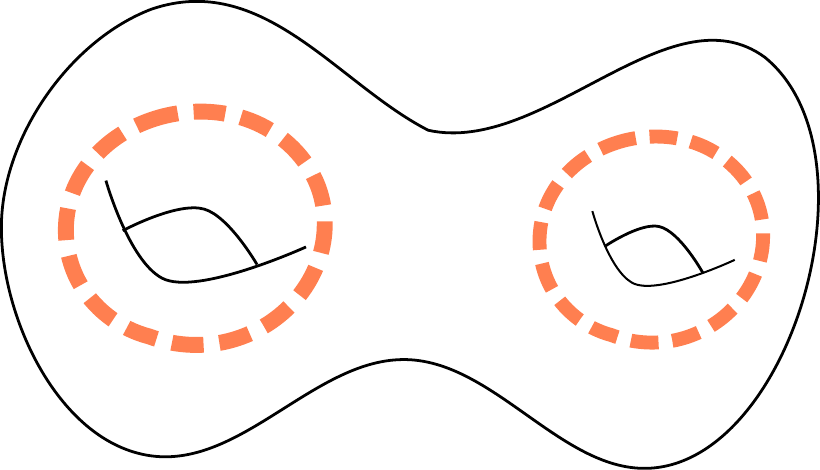}
\end{center}
\caption{The UV curve $\tilde{\CC}''_{2,0}$}
\end{subfigure}
\begin{subfigure}[b]{3in}
\begin{center}
 \includegraphics[width=2.5in]{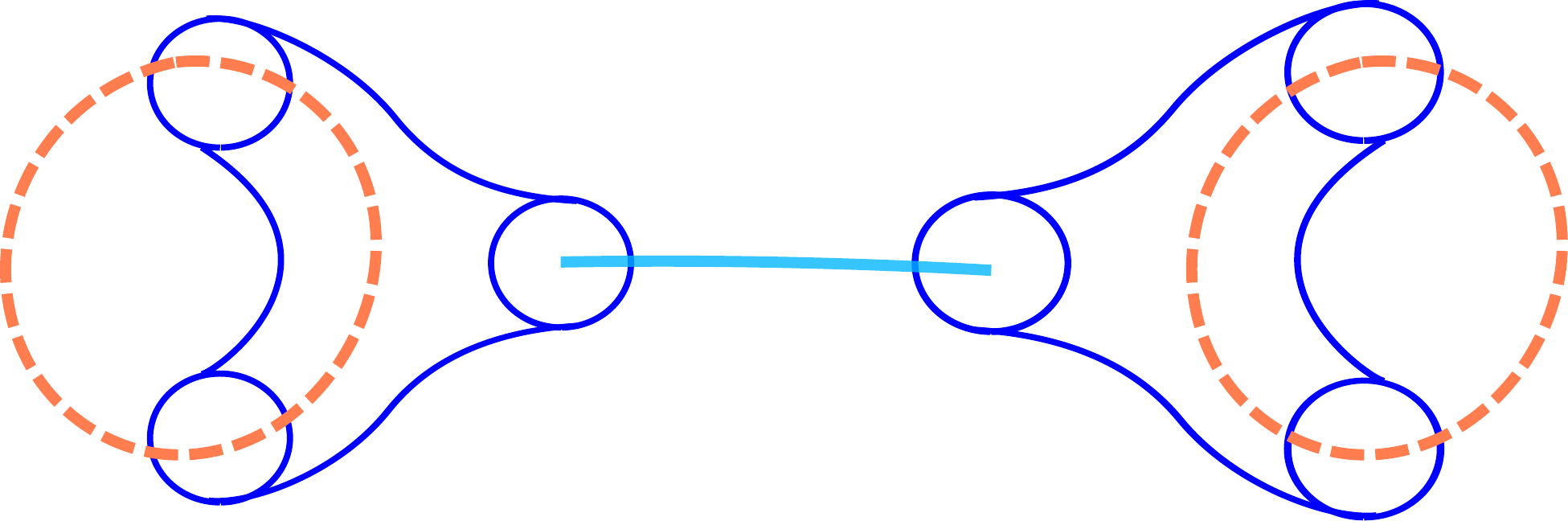}
 \end{center}
 \caption{A pair-of-pants decomposition.}
 \end{subfigure}
 \end{center}
 \caption{The UV curve $\tilde{\CC}_{2, 0}''$ and its pair-of-pants decomposition. There are two separate twist loops wrapping each holes. }
 \label{fig:wrap2hole2}
\end{figure}

Another dual frame is given as in the figure \ref{fig:wrap2hole2}. As we can see from the right-hand-side of the figure, this configuration can be obtained from taking the basic S-dual (crossing symmetry) of the middle node of the figure \ref{wrap2hole}. This can be thought of as having two twist loops wrapping each holes, leading us to different topology of the twist line configuration. 
Therefore, we conclude that there is only one physically inequivalent configuration of the twist line in a sense that all the  configurations are related by a chain of dualities.

\paragraph{S-dualities in the presence of the twist lines} 

\begin{figure}[!h]
\centering
\includegraphics[width=0.8\textwidth]{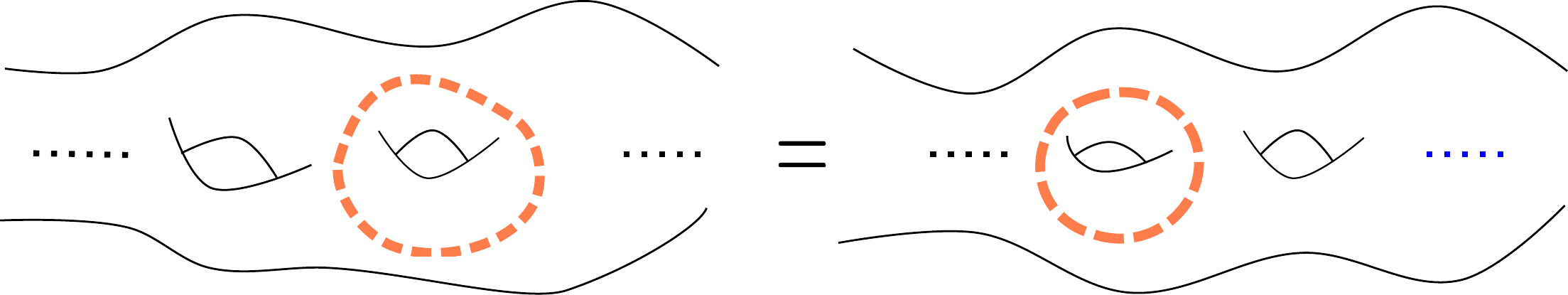} \\
\vspace{0.2in}
\includegraphics[width=0.45\textwidth]{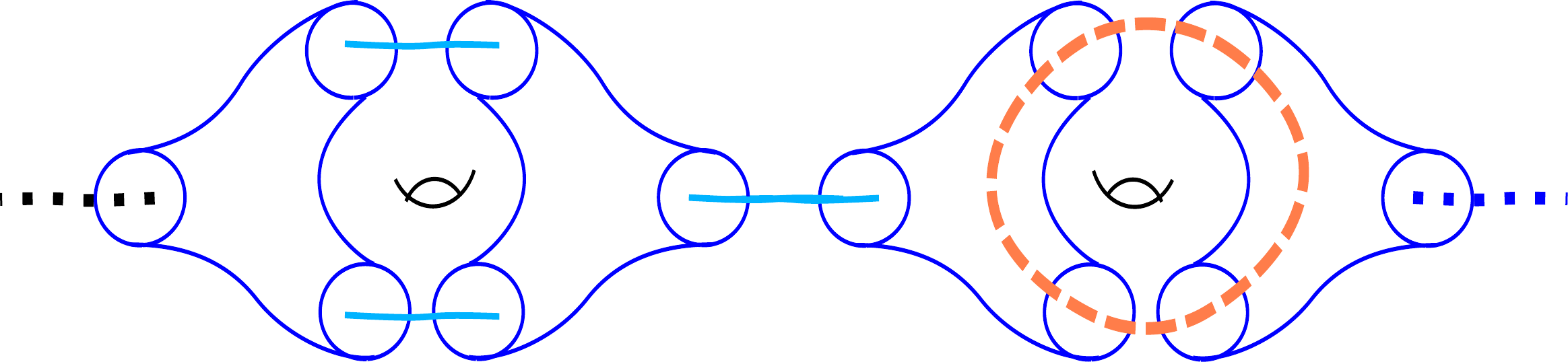}
\hspace{0.3in}
\includegraphics[width=0.45\textwidth]{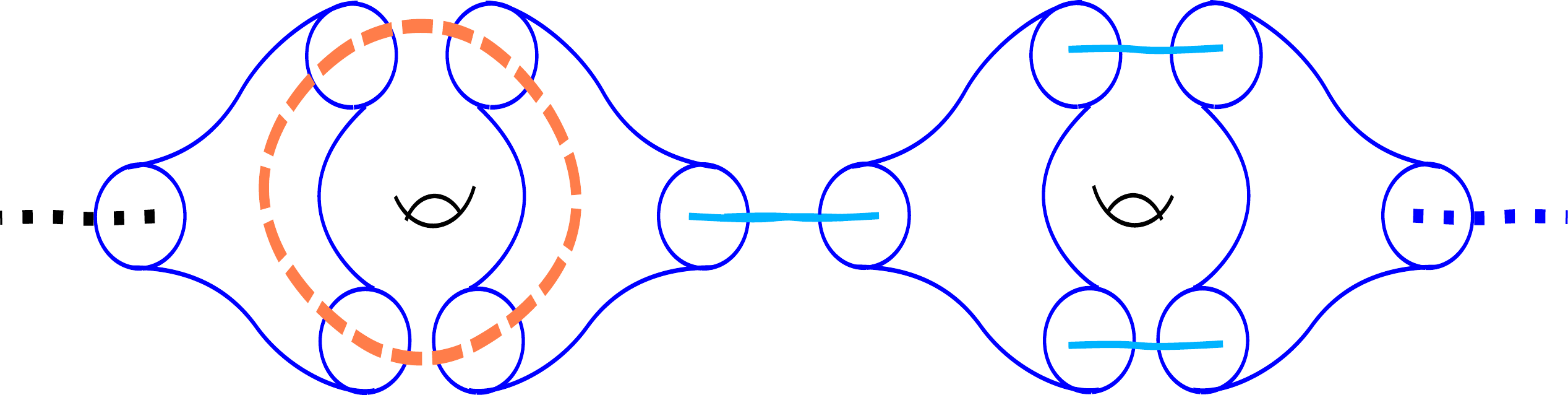}
\caption{Swapping holes and the pair-of-pants decompositions of corresponding configurations.}
\label{swapholescopy}
\end{figure}

\begin{figure}[!h]
\centering
\includegraphics[width=0.8\textwidth]{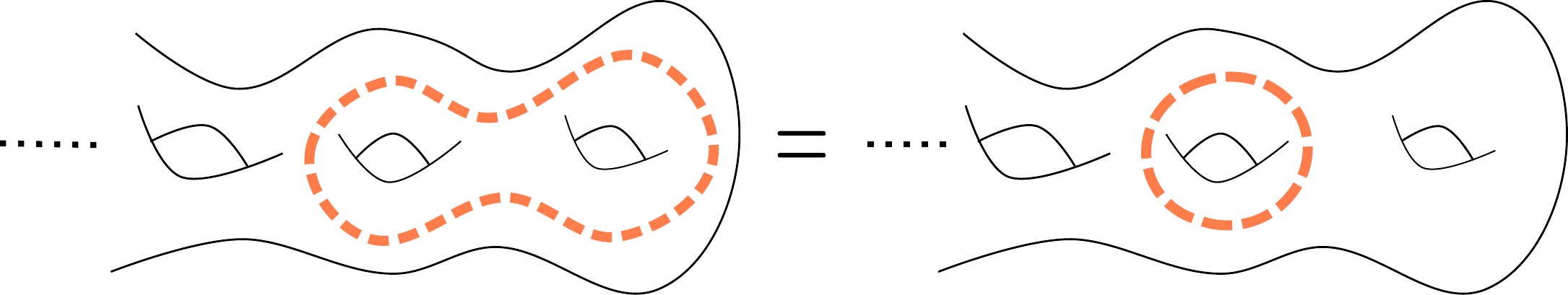} \\
\vspace{0.1in}
\begin{tabular}{cc}
  	\begin{minipage}[c]{0.45\textwidth}
  		\centering
		\includegraphics[width=0.8\textwidth]{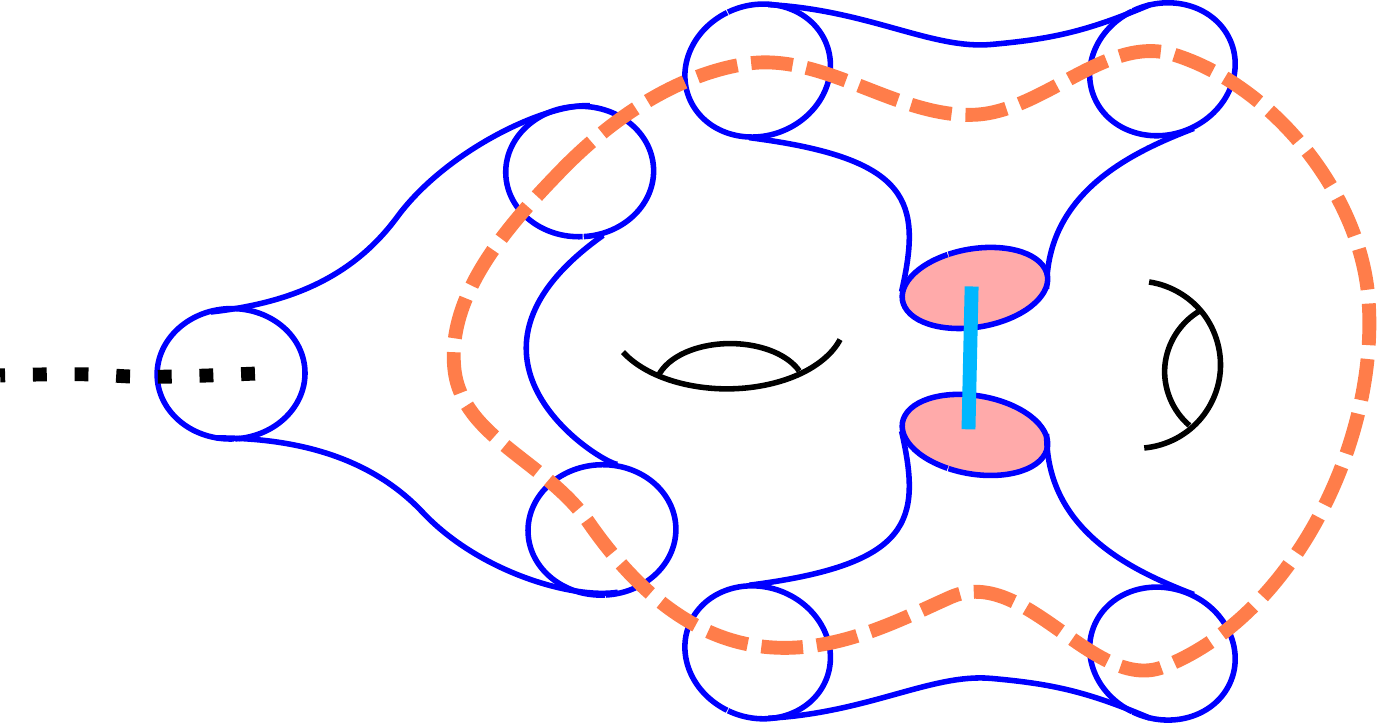}
	\end{minipage} & 
	\begin{minipage}[c]{0.45\textwidth}
	\centering
	\includegraphics[width=0.75\textwidth]{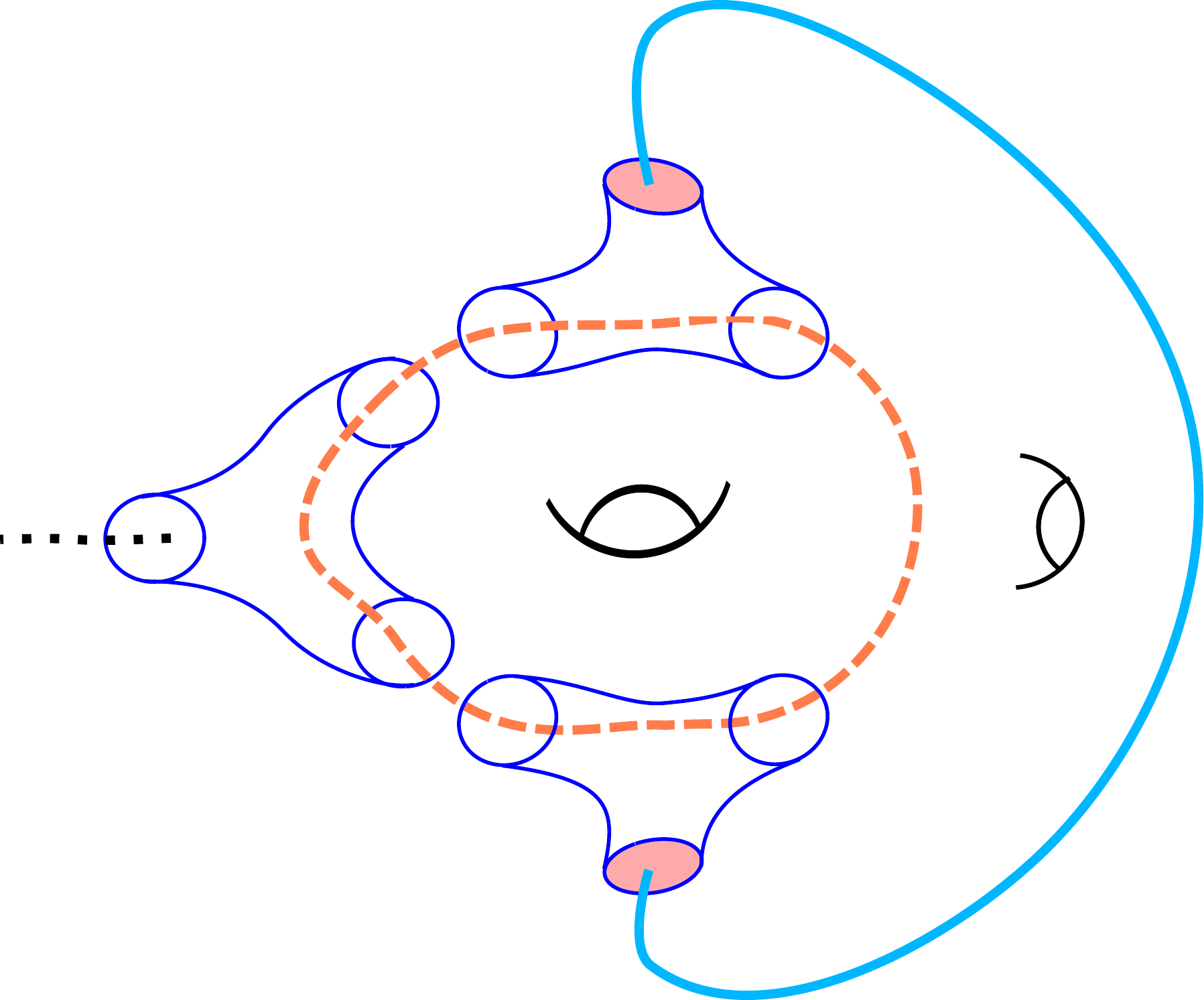}
	\end{minipage}
\end{tabular}
\caption{Upon decomposing the left configuration and gluing back upon flipping the pairs-of-pants, we obtain the configuration on the right, where twist loop wraps only one 1-cycle. }
\label{2=1}
\end{figure}

Now, let us show in general that the different configurations for the twist lines are dual to one another. 
The configurations for the twist loops are related to one another by the following operations:
\begin{enumerate}
\item swapping the 1-cycles as in figure \ref{swapholescopy}. 
\item one twist line wrapping two cycles can be replaced by the one twist line wrapping one hole as in figure \ref{2=1}.
\end{enumerate}
All the other configurations for the twist loops can be eventually mapped to the one wrapping a single 1-cycle upon performing sequence of two operations. 
We see that the first operation of swapping the holes can be done my deforming the geometry in a continuous way. Therefore the two configuration belongs to the same conformal manifold. So there is no preferred choice of pair-of-pants decomposition among the two. In other words, one can obtain two different pair-of-pants decompositions related by duality. 

The second operation can be justified by looking at the pair-of-pants decompositions. There is a dual frame where we obtain the same theory upon flipping the pairs-of-pants. One can obtain the other side of the twist line configuration by 1) decomposing into the pair-of-pants, 2) rotating some of the pairs-of-pants, 3) glue them back in such a way that the twist line wraps around only one cycle. See the figure \ref{2=1}.
Therefore, for the UV curve with some configuration of twist loops, we can always pick a dual frame which involves only one twist loop wrapping one 1-cycle on the curve. 

Notice that the first operation can be obtained by a successive application of the second operation. First, start with the twist loop wrapping one hole. It is now identical to the configuration where single twist line wrapping two neighboring holes. Finally, one can `shrink' the loop by choosing the other hole. 

\begin{figure}[!h]
\centering
\includegraphics[width=0.8\textwidth]{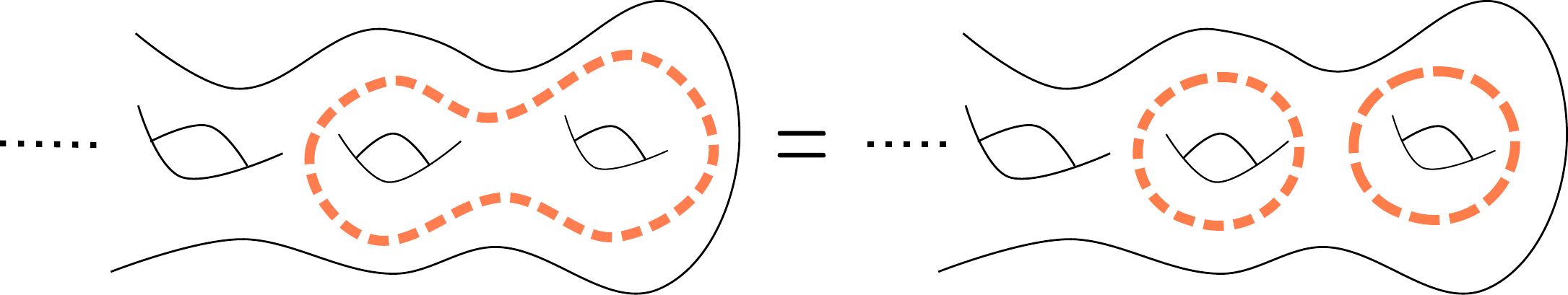} \\
\vspace{0.1in}
\includegraphics[width=0.3\textwidth]{2=1popL}
\hspace{0.4in}
\includegraphics[width=0.45\textwidth]{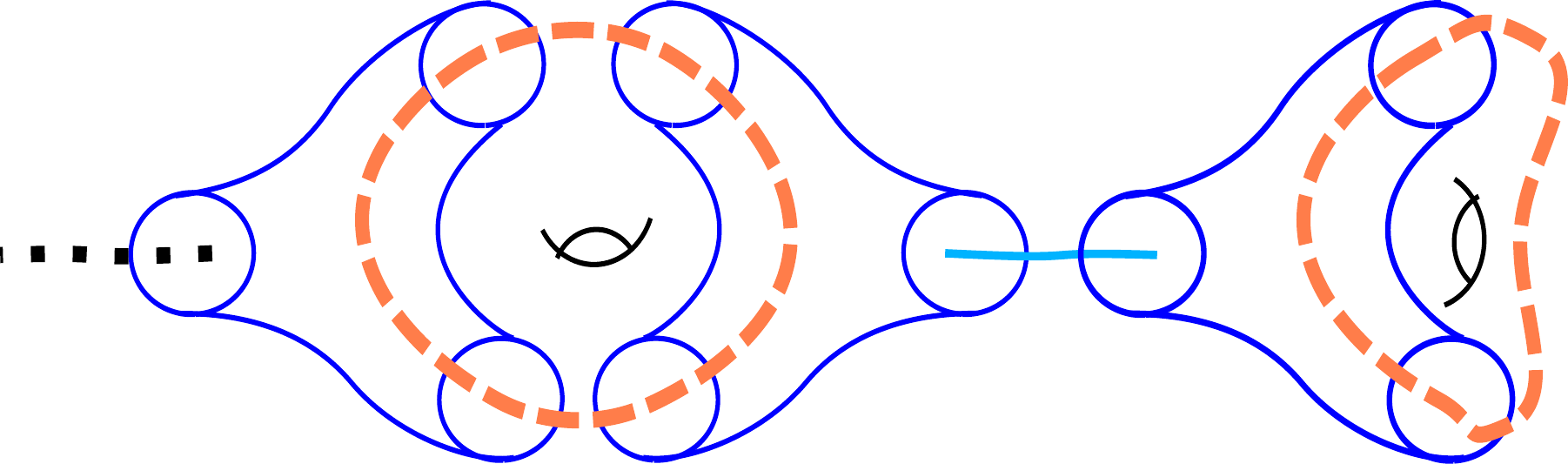}
\caption{Joining two twist loops to obtain one loop wrapping around two holes and pair-of-pants decompositions of two configurations. }
\label{joinholes}
\end{figure}
Let us point out that the equivalence of the configuration in figure \ref{joinholes} can also be obtained by performing duality (crossing symmetry) for the basic block of the 4-punctured sphere. Let us note that from the perspective of the definition of the twist line on the curve, the equivalence is a simple fact coming from the property of the branch cuts. But from the field theory perspective, this equivalence leads to the duality.


\subsection{Adding punctures} \label{sec:dualPunctures}
Until now, we have been considering the UV curve with no punctures i.e $\mathcal{C}_{g,0}$. When there is no puncture, the theory has no non-R global symmetry (for the $\CN=1$ case we discuss later in this section, there is only $U(1)$ non-R symmetry). By adding punctures to the UV curve, we obtain theories with larger global symmetries. 

\begin{figure}[!h]
\begin{center}
 \includegraphics[width=0.6\textwidth]{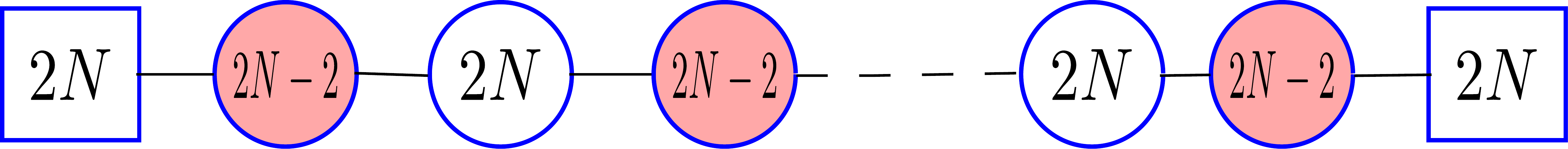}
 \end{center}
 \vspace{0.1in}
 \begin{center}
  \includegraphics[width=0.7\textwidth]{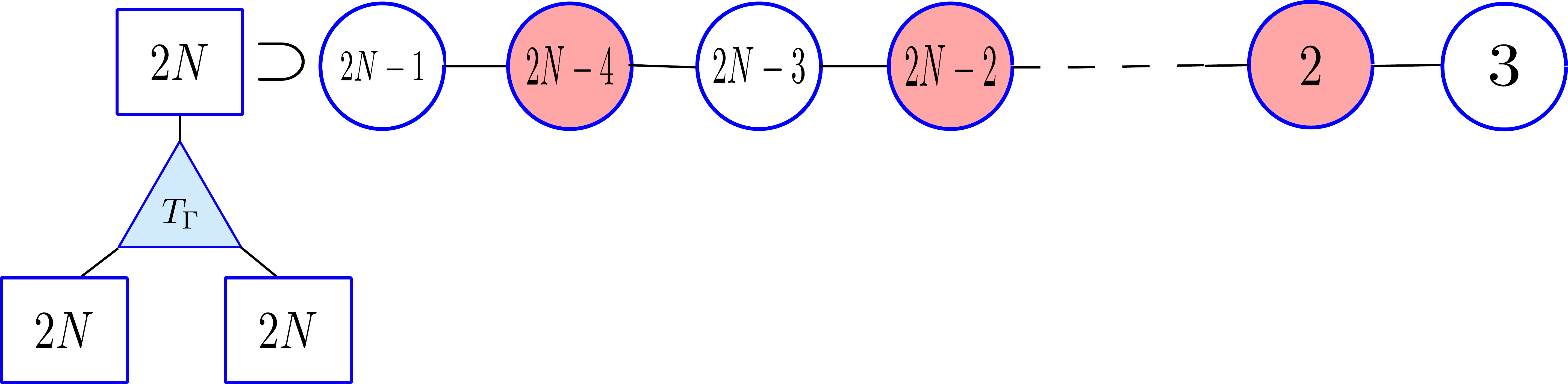}
 \end{center}
 \caption{Theory A: A linear quiver with $SO(2N)\times SO(2N)$ flavor symmetry and its dual where we have $T_{SO(2N)}$ with a superconformal tail. The shaded nodes are $USp$ groups.}
\label{fig:linearquiver1}
\end{figure}
\begin{figure}[!h]
\begin{center}
\includegraphics[width=0.6\textwidth]{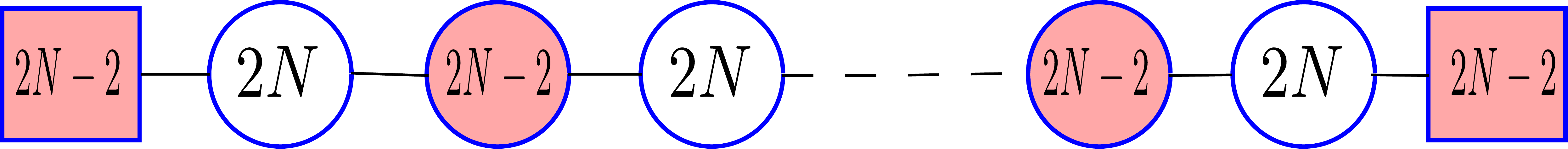}
\end{center}
\vspace{0.1in}
\begin{center}
\includegraphics[width=0.7\textwidth]{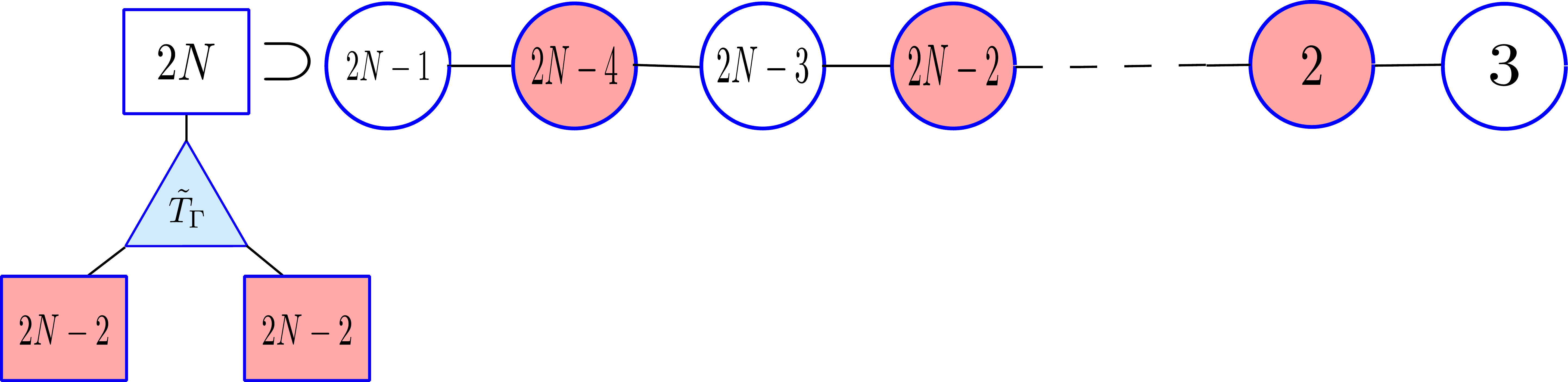}
\end{center}
\caption{Theory B: A linear quiver with $USp(2N-2)\times USp(2N-2)$ flavor symmetry and its dual where we have $\tilde{T}_{SO(2N)}$ with a superconformal tail. The shaded nodes are $USp$ groups.}
\label{fig:linearquiver2}
\end{figure}

Let us consider a linear quiver theory, which can be obtained by a sphere with a number of punctures. It is known \cite{Agarwal:2013uga} that linear $SO-USp$ quiver has a dual frame yields $T_{SO(2N)}$ or $\tilde{T}_{SO(2N)}$ block with the superconformal tail depending on whether the quiver has $SO(2N)$ or $USp(2N-2)$ flavor symmetry at the ends. The dualities are given as in the figure \ref{fig:linearquiver1} and figure \ref{fig:linearquiver2}. Let us call them theory A and B respectively. 
\begin{figure}[!h]
\begin{subfigure}[c]{\textwidth}
\begin{center}
\includegraphics[width=0.7\textwidth]{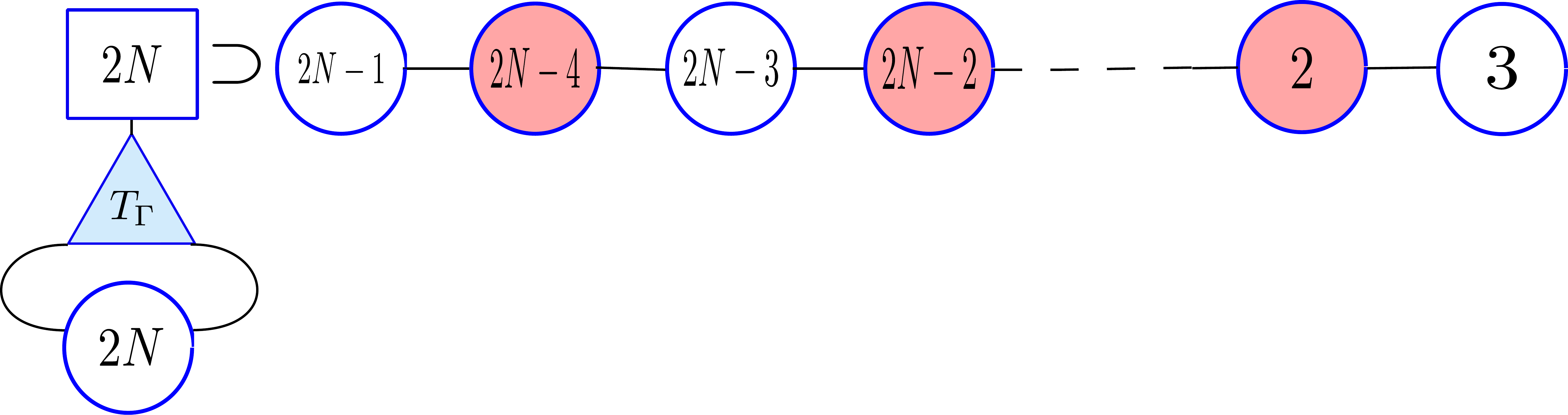}
\end{center}
\caption{Theory C formed by gluing the two ends of the quiver A.}
\end{subfigure}
\begin{subfigure}[c]{\textwidth}
\begin{center}
\vspace{0.2in}
\includegraphics[width=0.7\textwidth]{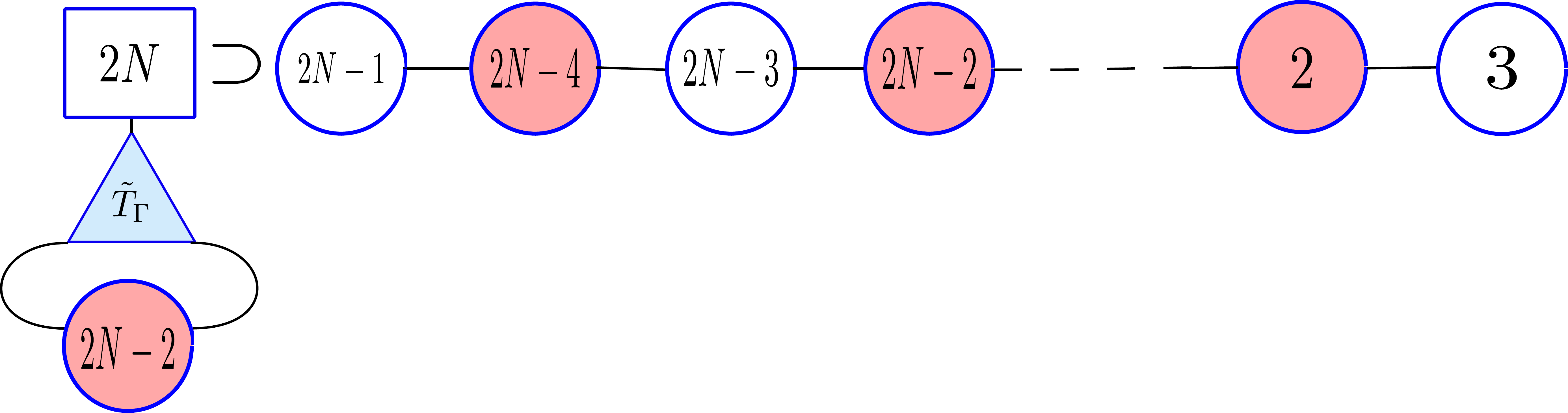}
\end{center}
\caption{Theory D formed by gluing the two ends of the quiver B.}
\end{subfigure}
\caption{Theories C and D obtained from theory A and B as in figure \ref{fig:linearquiver1}, \ref{fig:linearquiver2} upon gluing flavor symmetry of $T_{SO(2N)}$ and $\tilde{T}_{SO(2N)}$ respectively. They are dual to the circular quiver of figure \ref{circularquiver}. The shaded nodes denote the $USp$ groups.}
\label{gluedT}
\end{figure}

\begin{figure}[!h]
\begin{center}
\includegraphics[width=0.4\textwidth]{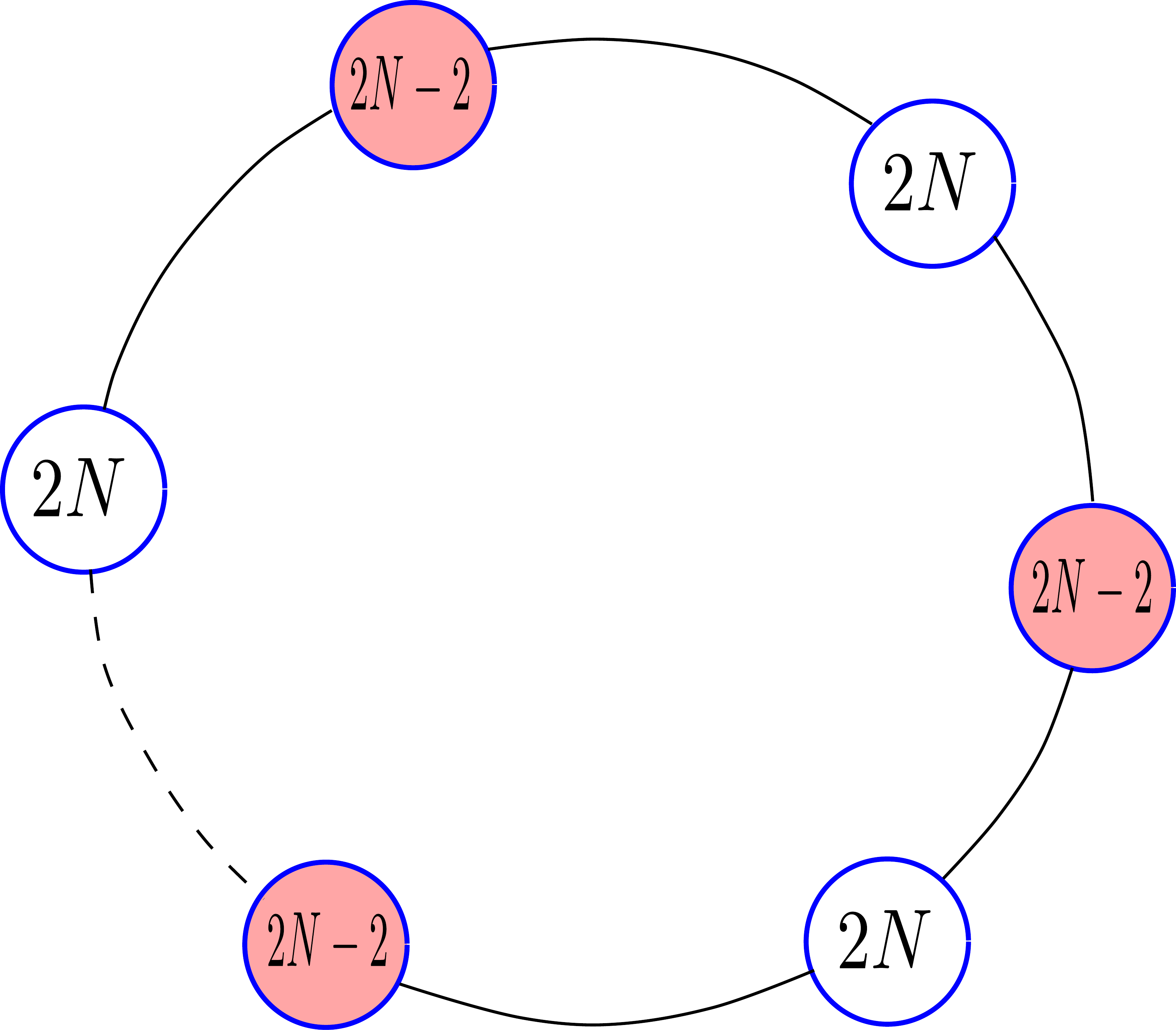}
\end{center}
\caption{Circular $SO-USp$ quiver. The shaded nodes are $USp$ groups.}
\label{circularquiver}
\end{figure}

Now in each of the dual frames, we gauge the flavor symmetry of $T_{SO(2N)}$ or $\tilde{T}_{SO(2N)}$ block to obtain the theory $C$ and $D$ respectively as in the figure \ref{gluedT}. 
This amounts to gluing of the flavor symmetry of the linear quiver, which results in the formation of a circular quiver. Now, it is easy to see that upon gluing, both $A$ and $B$ yields the same circular quiver as in figure \ref{circularquiver}. This implies that theory $C$ and $D$ are dual to each other and also to the the circular quiver.

\begin{figure}[!h]
\centering
\begin{subfigure}[t]{0.49\textwidth}
\centering
 \includegraphics[width=0.8\textwidth]{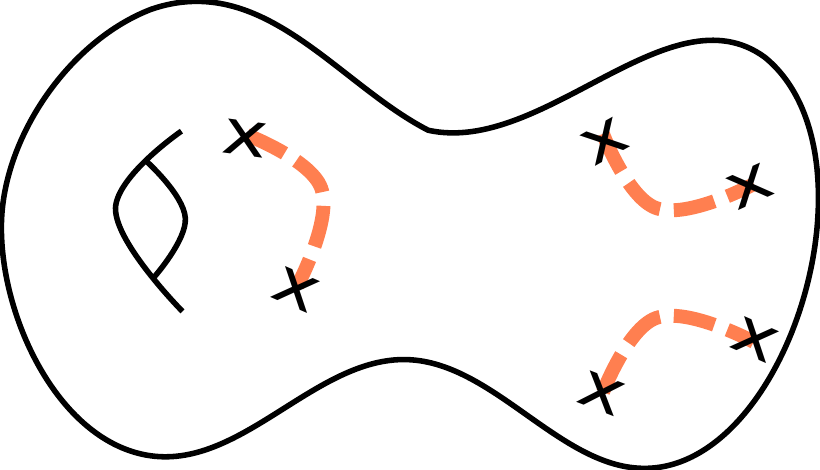}
 \caption{The UV curve $\CC_{1,6}$ with $6$ twisted punctures. This corresponds to the theory $C$ as in figure \ref{gluedT}.}
 \end{subfigure}~~~~
\begin{subfigure}[t]{0.49\textwidth}
  \centering
  \includegraphics[width=0.8\textwidth]{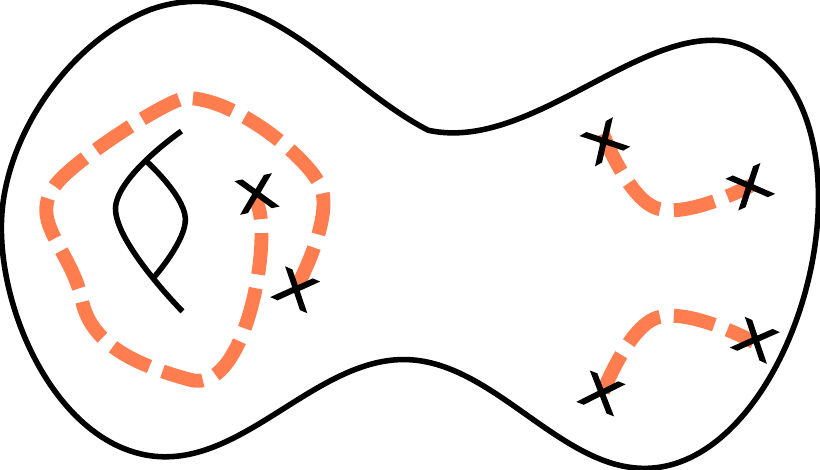}
 \caption{Move the twisted punctures to wrap around the hole. }
 \vspace{0.2in} 
 \end{subfigure}
\begin{subfigure}[t]{0.49 \textwidth}
\begin{center}
\includegraphics[width=0.8\textwidth]{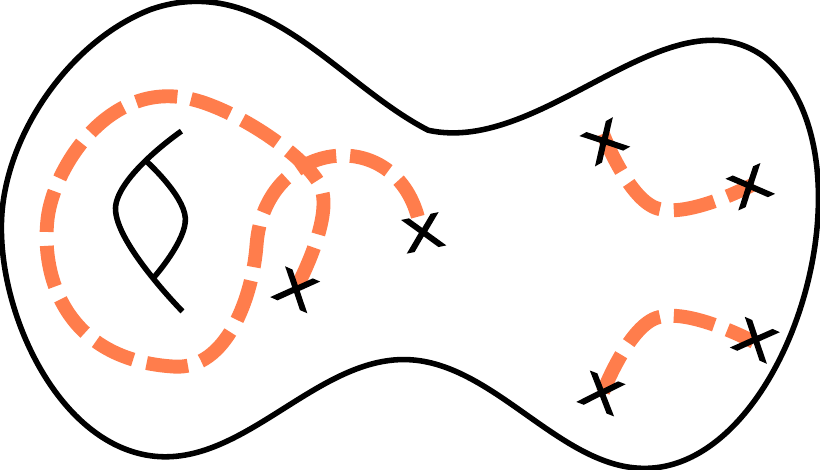}
\end{center}
\caption{The UV curve $\tilde{\CC}_{1,6}$, where the twist line wraps the hole as the punctures cross each other without colliding.}
\end{subfigure}~~~~
\begin{subfigure}[t]{0.49\textwidth}
\begin{center}
\includegraphics[width=0.8\textwidth]{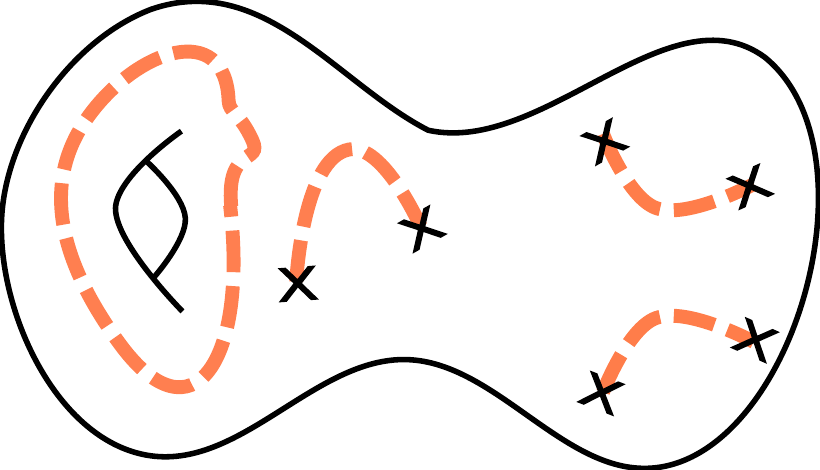}
\end{center}
\caption{Snip a part of the twist line, resulting the UV curve $\tilde{\CC}_{1,6}$ with an extra twist loop. This corresponds to the theory $D$ in figure \ref{gluedT}.}
\end{subfigure}
\caption{Construction of the UV curve with a twist loop wrapping a hole from the UV curve without the twist loop.}
\label{fig:oneholewrap}
\end{figure}
We can explain this duality in terms of the UV curve and twist lines as in figure \ref{fig:oneholewrap}. The UV curve $\CC_{1,6}$ corresponds to theory $C$, where we have $T_{SO(2N)}$ block with two of its $SO$ flavor symmetry gauged, once the punctures are chosen to be the twisted null puncture. The UV curve $\tilde{\CC}_{1,6}$ corresponds to theory $D$ with a $\tilde{T}_{SO(2N)}$ block and the $USp$ flavor symmetry gauged. 

Here, we see that the existence of the twist line wrapping around a loop in the UV curve does not change the theory, as long as there is a pair of twisted puncture in the UV curve already. One can have this extra twist loop from moving the twisted punctures around as described in the figure \ref{fig:oneholewrap}. 

Let us note that for $\G=A_{n}, E_{6}$ we do not have a linear quiver description as in the case of $D_n$. Nevertheless the argument in terms of the UV curve and the twist line should go through. Therefore similar construction should yield new dualities for all $\G=A_{n},D_{n},E_{6}$, regardless of the type of punctures as long as they are twisted. Therefore we arrive at the statement that 
\begin{align}
 \CT[\CC_{g, n}] \simeq \CT[\tilde{\CC}_{g, n}] \qquad \textrm{(if any of the punctures are twisted)} \ , 
\end{align}
where $\CC_{g, n}$ and $\tilde{\CC}_{g, n}$ are identical UV curves but the latter one has a twist loop.


\subsection{$\CN=1$ dualities}
\paragraph{$\CN=1$ class $\CS$ theory}
Let us generalize our discussion to the $\CN=1$ class $\CS$ theories \cite{Bah:2012dg, Gadde:2013fma,Xie:2013gma, Bah:2013aha, Agarwal:2013uga,Agarwal:2014rua,Agarwal:2015vla}. 
In addition to the UV curve we choose, an $\mathcal{N}=1$ class $\mathcal{S}$ theory is specified by the choice of the normal bundles $\CL(p)\oplus \CL(q) \to \CC_{g, n}$ over the UV curve. The two integers $(p, q)$ denote the degrees of the line bundles. In order to preserve any amount of supersymmetry, they have to satisfy $p+q=2g-2+n$ for the UV curve having genus $g$ and $n$ punctures. When one of $p$ or $q$ is set to zero, we obtain $\mathcal{N}=2$ theory. To each puncture, we also assign $\IZ_2$-valued color $\s=\pm$. 
These theories admit $U(1)_{+} \times U(1)_{-}$ global symmetry generated by what we call $(J_{+},J_{-})$. It comes from a subgroup of $SO(5)_R$ symmetry of the 6d $\CN=(2,0)$ theory, which is preserved after the partial topological twist on the UV curve. 

Similar to the $\CN=2$ case, we consider pair-of-pants decompositions of $\CC_{g,n}$ to obtain various dual descriptions of the SCFT. However, now to each pair-of-pants, we assign a color $\sigma=\pm$.\footnote{Generally, it is labelled by $(p, q) \in \IZ$ with $p+q=1$. The choice $\pm$ refers to $(p, q)=(1, 0)$ and $(0, 1)$ respectively. Other choices of $(p, q)$ are possible, but we do not discuss them here. See \cite{Agarwal:2015vla} for more detail.} The numbers of $\pm$-colored pairs-of-pants are given by the degrees of the line bundles $p$ and $q$ respectively. To each $\s$-colored pair-of-pants, we associate the $T_\G$ theory if there is no twist line. If there is a twist line running through the pair-of-pants, we associate the $\tilde{T}_\G$ instead. The color on the pair-of-pants, does not change the theory $T_\G$ or $\tilde{T}_\G$ itself. Instead, it determines how the gluing should be done. 

The gluing is done in the following manner: if $\sigma_{i}=\sigma_{j}$, the $i$-th and $j$-th pairs-of-pants are glued with the $\mathcal{N}=2$ vector multiplet. If $\sigma_{i}\neq\sigma_{j}$, they are glued with the $\mathcal{N}=1$ vector. When the gluing is done using $\CN=2$ vector, we add the following superpotential
\begin{align}
 W_{\CN=2} = \tr \phi (\mu_i + \mu_j ) \, 
\end{align}
for each $\CN=2$ nodes, where $\mu_i$, $\mu_j$ are the moment maps associated to the punctures and $\phi$ is the adjoint chiral multiplet in the $\CN=2$ vector multiplet. For the $\CN=1$ nodes, we add
\begin{align}
 W_{\CN=1} = \tr \mu_i \mu_j 
\end{align}
instead. We preserve $U(1)_+ \times U(1)_-$ global symmetry upon this gluing, where the charges for the $\mu_i$ are $(J_+, J_-) = (1+\s_i, 1-\s_i)$ and for the $\phi$ are $(J_+, J_-) = (1-\s_i, 1+\s_i)$. 

Finally, when the color of the puncture is identical to the pair-of-pants it belongs to, then we do nothing. But if the color of the puncture is opposite to that of the pair-of-pants, we add a chiral multiplet $M$ transforming under the adjoint of $\G$ (or $G$ if the puncture is twisted) and couple via superpotential $W= \tr M\mu$. 

The symmetries $U(1)_+$ and $U(1)_-$ both are the candidate $R$-symmetry of the theory so that the superpotential should have the charge $(J_+, J_-)=(2, 2)$. The superconformal $R$ symmetry can be written as
\begin{align}
 R_{IR} = \frac{1+\e}{2} J_+ + \frac{1-\e}{2} J_- = R_0 +\e \CF \ , 
\end{align}
where $R_0 = \half (J_+ + J_-)$ and $\CF = \half (J_+ - J_-)$. The value of $\e$ is fixed through the $a$-maximization \cite{Intriligator:2003jj}. 

\begin{figure}[!h]
\begin{subfigure}[c]{\textwidth}
\centering
\begin{tabular}{cc}
  \begin{minipage}[c]{0.48\textwidth}
  \centering
 \includegraphics[width=0.8\textwidth]{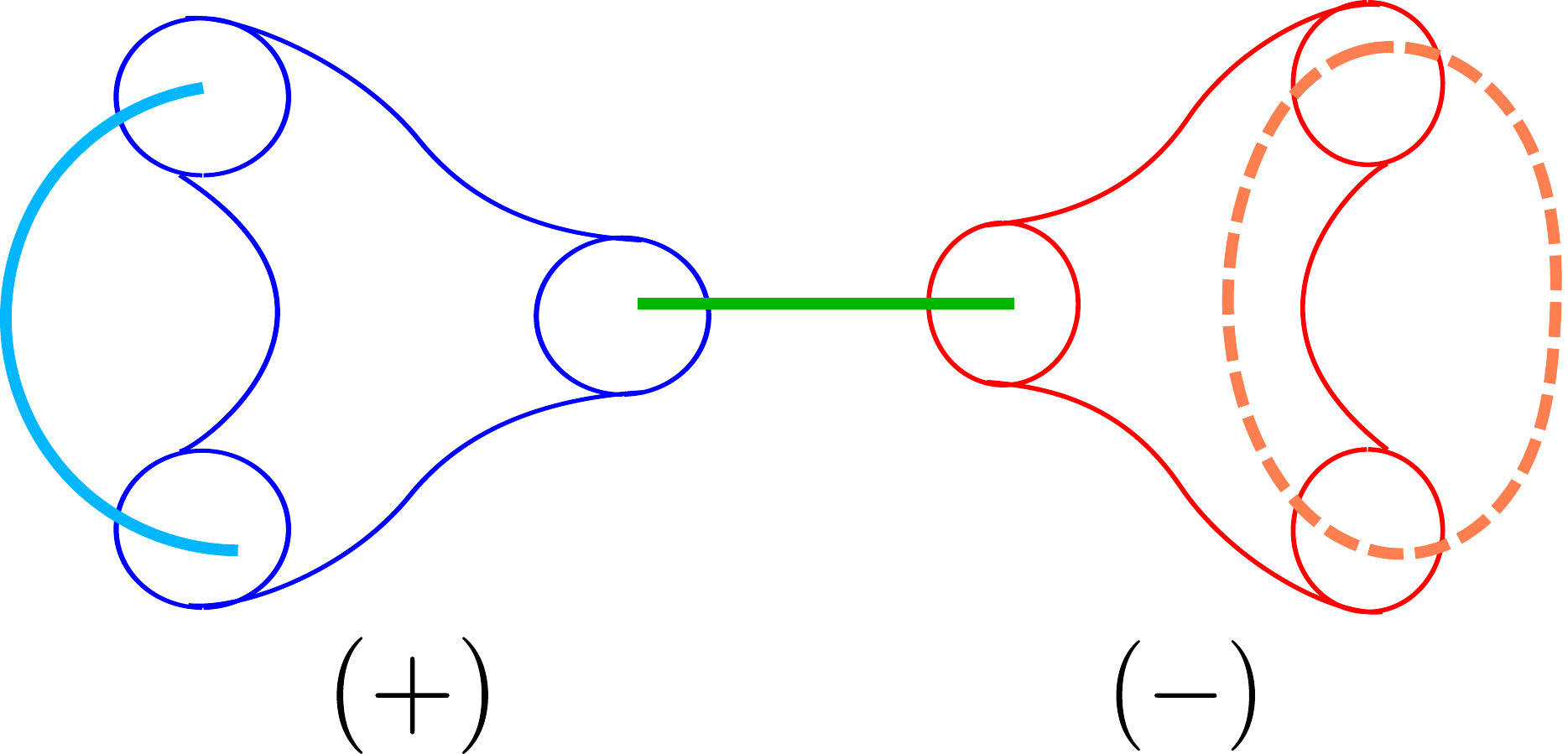}
 \end{minipage} &
  \begin{minipage}[c]{0.48\textwidth}
  \centering
  \includegraphics[width=0.9\textwidth]{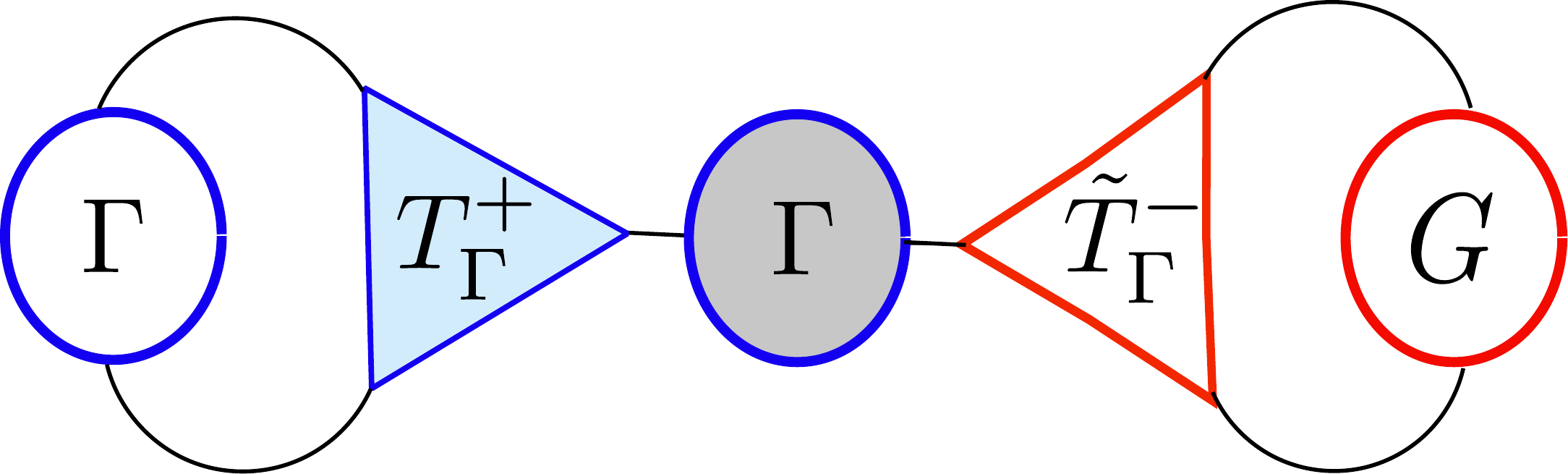}
  \end{minipage}
\end{tabular}
 \caption{A pair-of-pants decomposition.}
 \end{subfigure}
 
  \begin{subfigure}[c]{\textwidth}
  \begin{center}
  \vspace{0.1in}
  \begin{tabular}{cc}
 \includegraphics[height=0.28\textwidth]{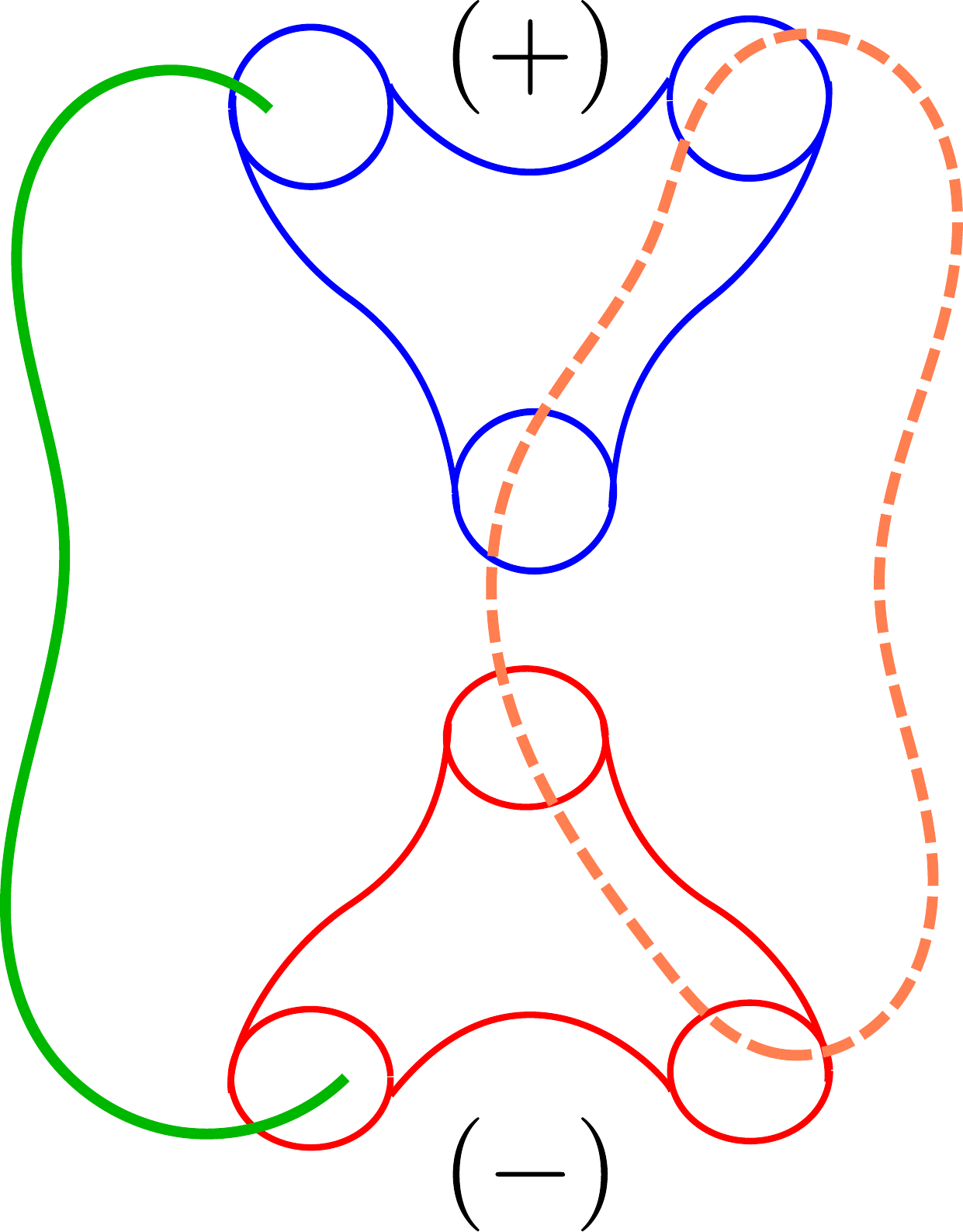} &
 \hspace{0.6in}
  \includegraphics[width=0.28\textwidth]{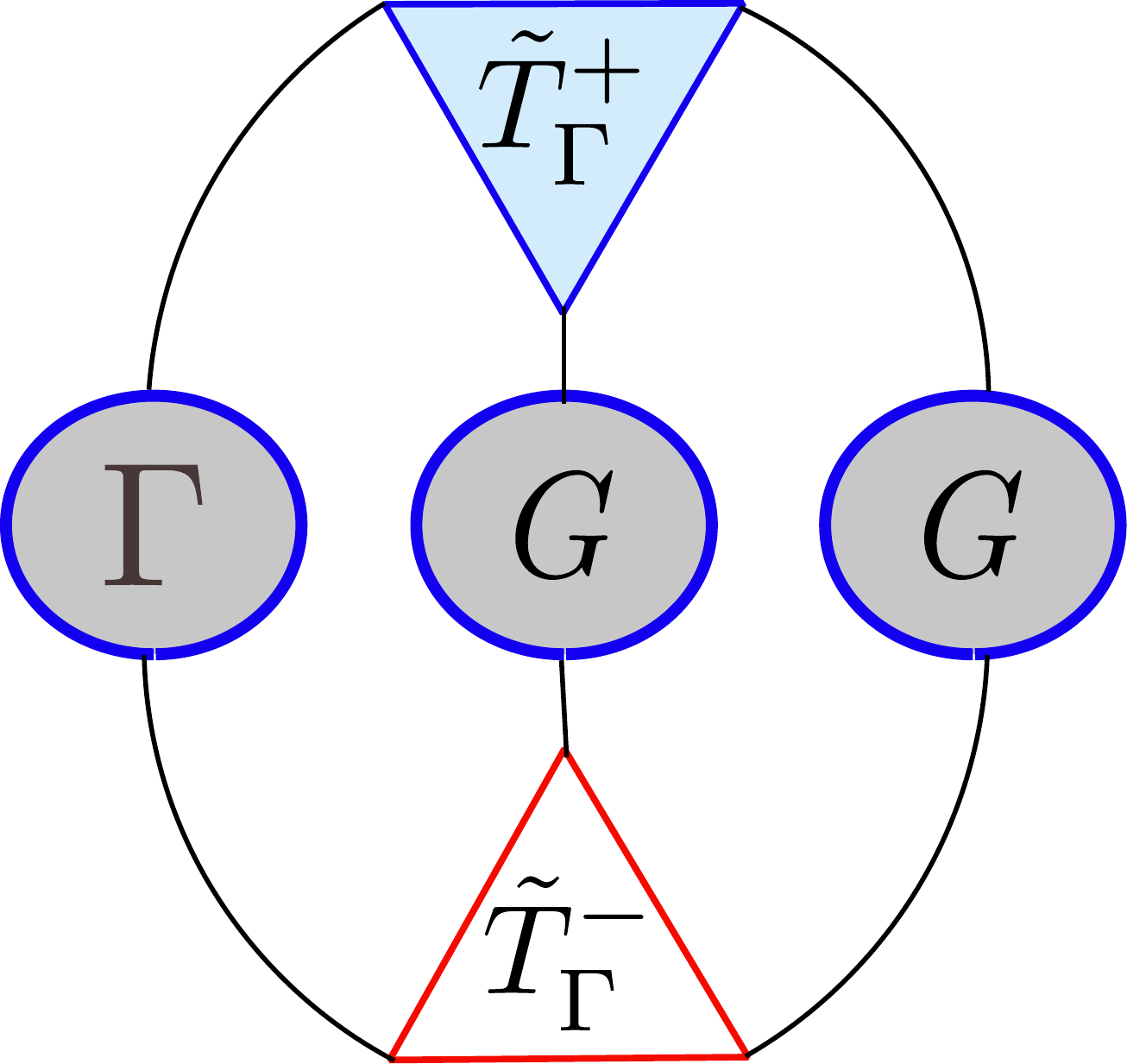}
  \end{tabular}
\end{center}
\caption{Another pair-of-pants decomposition.}
\end{subfigure}
\caption{Two ways of decomposing the UV curve $\tilde{\CC}_{2,0}$ (labelled as $a$ and $b$). The shaded node refers to $\CN=1$ gauging whereas unshaded node refers to $\CN=2$ gauging.}
\label{fig:onewholewrapN1}
\end{figure}

\paragraph{Dualities} 
\begin{figure}[!h]
\begin{center}
 \includegraphics[width=3.5cm]{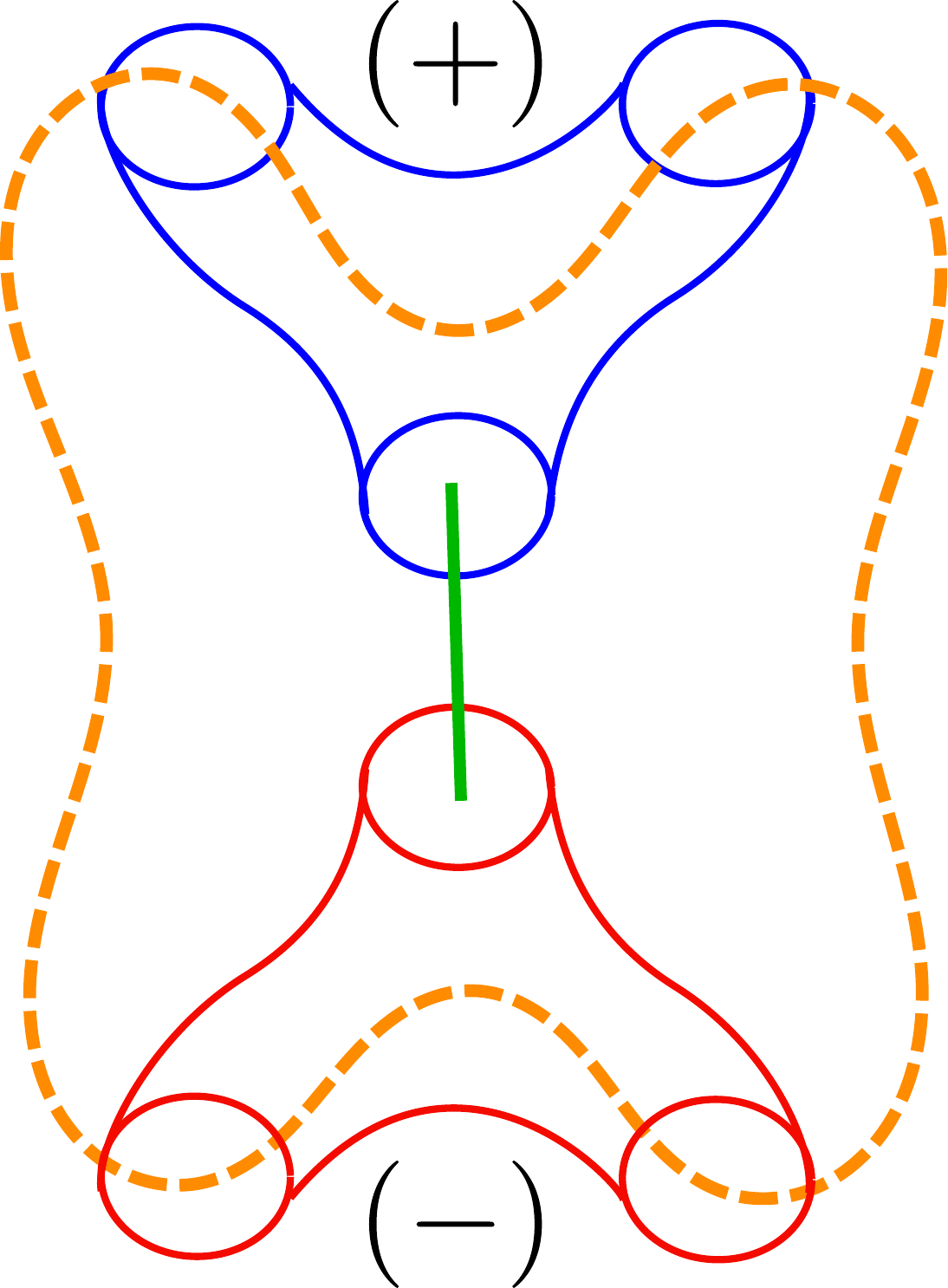}
\end{center}
\caption{The UV curve $\tilde{\CC}'_{2, 0}$. The twist line (orange) wraps around two holes. The pair-of-pants decomposition reveals that we get the same theory as the case where the twist line wraps only one hole.}
\label{fig:twoholewrapN1}
\end{figure}

Let us consider an example: Choose the UV curve as $\CL(1) \oplus \CL(1) \to \tilde{\CC}_{2,0}$ with the twist line wrapping one of the holes. Two pair-of-pants decompositions will yield two dual frames for the same theory. See the figure \ref{fig:onewholewrapN1}. 
The first one will yield a $T_\G$ block and a $\tilde{T}_\G$ block with $\G_{\CN=2} \times \G_{\CN=2} \times G_{\CN=1}$ gauge symmetry glued with $\CN=1$ and $\CN=2$ vector multiplets, while the second decomposition will yield two $\tilde{T}_\G$ blocks with $ \G_{\CN=1} \times G_{\CN=1} \times G_{\CN=1}$ gauge symmetry all $\CN=1$ vectors. 

Just as in the case of $\mathcal{N}=2$, we can consider another UV curve $\tilde{\CC}'_{2,0}$, where the twist line wraps around two holes of the UV curve. 
The pair-of-pants decomposition as in the figure \ref{fig:twoholewrapN1} reveals that it yields same theory as $\tilde{\CC}_{2,0}$ even though $\tilde{\CC}_{2,0}$ is topologically distinct from $\tilde{\CC}'_{2,0}$. Therefore, there is only one physically inequivalent configuration for the twist line. 

As we have seen in the case of $\CN=2$ theories, we can argue that there is a unique physically inequivalent configuration of the twist line for arbitrary genus $g\ge2$. The coloring of the pair-of-pants have no effect on the proof of our statements about the twist loop, upon replacing S-dualities by IR dualities.


\section{Misleading anomalies and the superconformal indices} \label{sec:anomaly} 
In this section, we compute the 't Hooft anomalies and superconformal indices to verify the dualities. Also, we find that the theory $\CT[\CC_{g, n}]$ given by the UV curve without the twist loop and the theory $\CT[\tilde{\CC}_{g, n}]$ given by the same curve but with the twist loop share the same anomalies, but distinct indices. 

\subsection{Central charges of the $\CN=2$ theories} \label{sec:N2cc}
For any $4d$ $\CN=2$ SCFT, the conformal anomalies $(a, c)$ are determined in terms of the 't Hooft anomalies \cite{Kuzenko:1999pi}:
\begin{align}\label{anomalyN2}
 \tr R_{\CN=2} = \tr R_{\CN=2}^3 = 2 (n_v - n_h) \ , \qquad \tr R_{\CN=2} I_3^2 = \frac{1}{2}n_v \ , 
\end{align}
where $I_3$ and $R_{\CN=2}$ are the Cartans of the R-symmetry $SU(2)_R \times U(1)_r$. 
It is sometimes more convenient to use the notion of effective number of hypermultiplets and vector multiplets $n_h, n_v$ instead of using $a$ and $c$. They are related by
\begin{align}
 a = \frac{1}{24} n_h +\frac{5}{24} n_v \ , \qquad c = \frac{1}{12} n_h + \frac{1}{6} n_v \ . 
\end{align}
If the theory has a flavor symmetry, we can also define the flavor central charge $k$ as
\begin{align}
 k \delta^{ab} = -2 \tr R_{\CN=2} T^a T^b , 
\end{align}
where $T^a$ are the generators of the flavor symmetry. 

\begin{table}[t] 
\begin{center}
\begin{tabular}{|c|ccccccccc|}
\hline
 & $A_{n-1}$ & $B_n$ & $C_n$ & $D_n$ & $E_6$ & $E_7$ & $E_8$ & $F_4$ & $G_2$ \\
 \hline \hline
 $d$ & $n^2 - 1$ &  $n(2n+1)$ & $n(2n+1)$ & $n(2n-1)$ & $78$ & $129$ & $248$ & $52$ & $14$ \\
 $h^\vee$ & $n$ & $2n-1$ & $n+1$ & $2n-2$ & $12$ & $18$ & $30$ & $9$ & $4$ \\
 \hline
\end{tabular}
\end{center}
\caption{The dimension $d$ and the dual Coxeter number $h^\vee$ for all simple Lie groups.}
\label{table:group}
\end{table}
The central charges for the $T_\G$ theory can be computed by adding the local contribution from the punctures and the bulk contribution from the sphere \cite{Chacaltana:2012zy}. For each of the maximal punctures, we have
\begin{align}
 n_h = \frac{2}{3} h^\vee_\G d_\G \ , \qquad n_v = \frac{2}{3} h^\vee_\G d_\G + \half r_\G - \half d_\mathfrak{g} \ , 
\end{align}
where $h^\vee_\G$, $r_\G$ and $d_\G$ are the dual Coxeter number, rank and the dimension of $\G$ respectively and $d_\mathfrak{g}$ is the dimension of the $\mathfrak{g}$. If the puncture is untwisted, $\mathfrak{g}=\G$ and if it is twisted, it is given by an appropriate non-simply laced group $G$. They are summarized in the table \ref{table:group}. 
For the bulk of genus $g$, we get \cite{Benini:2009mz, Alday:2009qq}
\begin{align}
 n_h = (g-1) \left(\frac{4}{3} h^\vee_\G d_\G \right) \ , \qquad n_v = (g-1) \left( \frac{4}{3} h^\vee_\G d_\G + r_\G \right) \ . 
\end{align}
Adding up the contributions from the three maximal punctures and the $g=0$ bulk, we get the anomalies for the $T_\G$ theory as
\begin{align} \label{eq:Tac}
 n_v = \frac{2}{3} h^\vee_\G d_\G +\frac{1}{2}r_\G - \frac{3}{2}d_\G \ , \qquad n_h = \frac{2}{3} h^\vee_\G d_\G \ , \qquad k_\G =  2h^\vee_\G \ , 
\end{align}
where $k_\G$ is the flavor central charge for each of the three $\G$ symmetries. 

The $\tilde{T}_\G$ theory has two twisted punctures and one untwisted puncture, so that 
\begin{align} \label{eq:Ttwistedac}
 n_v = \frac{2}{3} h^\vee_\G d_\G +\frac{1}{2}r_\G - d_G - \frac{1}{2}d_\G \ , \quad n_h = \frac{2}{3} h^\vee_\G d_\G \ , \quad k_\G =  2h^\vee_\G \ , \quad k_G = 2h^\vee_G \ , 
\end{align}
where $k_G$ is the flavor central charge corresponding to the twisted punctures. Notice that the central charge formulae for the $\tilde{T}_\G$ are almost identical to that of the $T_\G$ theory except for the last term for $n_v$. 
For example, the number of effective vector multiplets $n_v$ and hypermultiplets $n_h$ for the $T_{SO(2N)}$ and $\tilde{T}_{SO(2N)}$ theory is given as follows \cite{Agarwal:2013uga}:
\begin{align} 
\begin{split}
n_{v}(T_{SO(2N)})& =\frac{1}{3}N(N-2)(8N-5) \ , \\
n_{v}(\tilde{T}_{SO(2N)})& = \frac{1}{3}(N-1)(8N^2 - 13N + 3)\ , \\
n_{h}(T_{SO(2N)}) &= n_{h}(\tilde{T}_{SO(2N)})= \frac{4}{3}N(N-1)(2N-1) \ . 
\end{split}
\end{align}
Since we know the central charges for the building blocks, one can obtain the central charges for the theories corresponding to the arbitrary UV curve of higher genus by adding each contribution from the pair-of-pants and the vector multiplets.

\paragraph{Central charges in the presence of the twist line}

Let us consider two Riemann surfaces $\CC_{g,n}$ and $\tilde{\CC}_{g,n}$, each with genus $g$ with $n$ punctures. Let us assume all the punctures are of the untwisted maximal type. The $\tilde{\CC}_{g, n}$ has a twist loop with the same geomettry as $\CC_{g, n}$. See the figure \ref{fig:genus3}, for example. 
Now we perform a pair-of-pants decomposition of $\CC_{g, n}$ and $\tilde{\CC}_{g, n}$ and write down the theory in term of $T_\G$ and $\tilde{T}_\G$, where they are glued with the $\mathcal{N}=2$ vector multiplets. (See the figure \ref{fig:g3PoP}.)
For the surface $\CC_{g, n}$ without twist loop, we have $2g-2+n$ pairs-of-pants, each of them representing $T_\G$ block and $3g-3+n$ of the gauge nodes with group $\G$. 

For the $\tilde{\CC}_{g, n}$ with a twist loop, some of the $T_\G$ blocks are replaced by the $\tilde{T}_\G$ block and also some of the gauge groups will be replaced by $G$. The number of the $\tilde{T}_G$ blocks depends on the choice of the dual frames, but when ever there is a twist line flowing through a pair-of-pants, it should replace the gauge group $\G$ by $G$. Also this is the case since we assume that the twist line is forming a loop. 
Therefore, the effective number of vector multiplets is given by 
\begin{align}
n_{v}= (2g-2+n) n_v (T_\G) +(3g-3+n) d_\G
 +  2 \alpha \left[ n_v(\tilde{T}_\G) - n_v(T_\G) + d_G- d_\G \right] \ , 
\end{align}
where $\a$ depends on the dual frames. When $\a=0$, it describes the case without the twist loop. 
Now, from \eqref{eq:Tac} and \eqref{eq:Ttwistedac}, we have
\begin{align} \label{eq:nvDiff}
\left[ n_v(\tilde{T}_\G)-n_v(T_\G) +d_G - d_\G \right]=0 \ .
\end{align}
Hence, the presence of the twist line or the choice of the dual frame does not affect the effective number of vector multiplets. 
Similarly, the effective number of hypermultiplets are given by
\begin{align} \label{eq:nhDiff}
n_{h}= (2g-2+n) n_h(T_\G) + 2 \alpha \left[ n_h(\tilde{T}_\G) -n_h(T_\G) \right] = (2g-2+n) n_h(T_\G) \ . 
\end{align}
Therefore, a twist loop does not alter the effective number of hypermultiplets as well. 
To summarize, the central charges $a$ and $c$ are unaltered even if we add a twist loop to the UV curve, namely the theories associated to the UV curves $\CC_{g,n}$ and $\tilde{\CC}_{g,n}$ have the same anomaly coefficients. Note that the flavor central charges all match as well, as long as the punctures are of the same type. 

The matching of anomalies is rather surprising, since the theories corresponding to each curve is not the same! One might think that this is a signal that they are dual to each other, but as we will see later, the superconformal indices for these two theories are different. 
The two theories having the identical anomaly coefficients have distinct superconformal indices if all the punctures are untwisted, so they are not in the same universality class. We also find that adding a twisted puncture makes the theory dual to each other as we have seen in section \ref{sec:dualPunctures}. For this case, the indices match as well. 

\paragraph{Coulomb branch operators}
Let us also mention that the theories corresponding to $\CC_{g, n}$ and $\tilde{\CC}_{g, n}$ have the identical Coulomb branch as well. 
The $T_{\G}$ and the $\tilde{T}_{\G}$ theories have their own Coulomb branches. One can obtain the dimensions of the operators parametrizing the Coulomb branch for the $\G=SO(2N)$ using the dualities described in the figure \ref{fig:linearquiver1} or \ref{fig:linearquiver2}. Both $T_{SO(2N)}$ and $\tilde{T}_{SO(2N)}$ theories have the Coulomb branch operators $u_{i, j}$ with the scaling dimensions 
\begin{align}
 [u_{i, j}] = 2i+2, \qquad & i=1, \ldots, N-2, \quad j=1, \ldots, 2i \ , \\ 
 [u_{N-1, j}] = N, \qquad & \begin{cases} j=1, \ldots N-2 \textrm{ for } T_{SO(2N)} \\
 							j=1, \ldots N-1 \textrm{ for } \tilde{T}_{SO(2N)} 	\end{cases} .
\end{align}
Therefore the Coulomb operators for the $T_{SO(2N)}$ and the $\tilde{T}_{SO(2N)}$ are almost identical except for the extra dimension $N$ operator in the $\tilde{T}_{SO(2N)}$ theory. Note that once we form a one-punctured torus by gauge the diagonal subgroup of $SO(2N)^2$ or $USp(2N-2)^2$ of the $T_{SO(2N)}$ or the $\tilde{T}_{SO(2N)}$ theory, we get the same set of Coulomb branch operators. One can easily verify that the set of Coulomb branch operators for the higher genus UV curve does not change upon including a loop of twist line. A combination of the conformal anomalies is determined by the dimensions of the Coulomb branch operators as \cite{Shapere:2008zf}
\begin{align}
 2a - c = \frac{1}{4} \sum_{i} ( 2[\CO_i] - 1) \ , 
\end{align}
where the sum is over all the Coulomb branch operators. Therefore, this combination of anomalies remain identical upon adding twist line, as we have already shown.

\subsection{Anomaly coefficients of the $\CN=1$ theories}
Now let us generalize our discussion to the $\CN=1$ theories. 
Any class $\CS$ theory preserves $U(1)_+ \times U(1)_-$ global symmetry whose generators we denote as $J_+$, $J_-$. Sometimes we use $U(1)_{R_0} \times U(1)_\CF$ where $R_0 = \half (J_+ + J_-)$ and $\CF = \half (J_+ - J_-)$. For the $\CN=2$ theory, they can be written in terms of $\CN=2$ $R$-charges as
\begin{align}
R_{0}=\frac{1}{2}R_{\CN=2}+I_{3}, \qquad \CF =  -\frac{1}{2}R_{\CN=2}+I_{3} \ . 
\end{align}
When constructing an $\CN=1$ theory by gluing the $T_\G$ or $\tilde{T}_\G$ blocks, we assign color $\s_i$ to each blocks and one linear combination of $U(1)_\CF$ for the blocks 
\begin{align}
 \CF = \sum_i \s_i \CF_i \ , 
\end{align}
is preserved upon gluing. Here $\CF_i$ is the $U(1)$ symmetry only acting on the $i$-th block. Let us write $T^\s_\G$ and $\tilde{T}^\s_\G$ to denote the coloring for each building blocks. 
Now, using the equation \eqref{anomalyN2}, the anomaly coefficients can be computed. The results are in the table \ref{table:N=1}. 
\begin{table}[h!] 
\begin{center}
\begin{tabular}{|c||c|c|c|c|}
\hline 
  & $\tr J_{+}=\tr J_{+}^{3}$ &  $\tr J_{-}= \tr J_{-}^{3}$ & $ \tr J_{+}J_{-}^{2}$ & $\tr J_{+}^{2} J_{-}$\\
 \hline \hline
 $T^{+}_\G$ &0 & $r_{\G}-3d_{\G}$ & 0& $\frac{4}{3} h^\vee_\G d_\G + r_\G - 3 d_\G$\\
 \hline
 $\tilde{T}^{+}_\G$&0 &$r_{\G}-d_{\G}-2d_{G}$  &0  &$\frac{4}{3} h^\vee_\G d_\G + r_\G - d_\G- 2d_{G}$\\
 \hline
 $T^{-}_\G$ & $r_{\G}-3d_{\G}$ &0 &$\frac{4}{3} h^\vee_\G d_\G + r_\G - 3d_\G$ &0\\
 \hline
 $\tilde{T}^{-}_\G$ &$r_{\G}-d_{\G}-2d_{G}$ &0 &$\frac{4}{3} h^\vee_\G d_\G + r_\G - d_\G-2 d_{G}$  &0\\
 \hline \hline
 $ \G^{+}$ &0 &$2d_{\G}$ &0  &$2d_{\G}$\\
 \hline
 $G^{+}$& 0& $2d_{G}$& 0 &$2d_{G}$ \\
 \hline
  $\G^{-}$ &$2d_{\G}$ & 0&$2d_{\G}$  &0\\
 \hline
   $G^{-}$& $2d_{G}$&0 &$2d_{G}$ &0\\
   \hline
   $\G$& $d_{\G}$ &$d_{\G}$ & $d_{\G}$&$d_{\G}$\\
   \hline
    $G$ & $d_{G}$&  $d_{G}$&$d_{G}$ &$d_{G}$\\
   \hline
\end{tabular}
\end{center}
\caption{Anomaly coefficients for the building blocks of the $\mathcal{N}=1$ class $\mathcal{S}$ theory. $\G$ and $G$ denotes the $\CN=1$ gauge nodes, and $\G^\s$, $G^\s$ denote the $\CN=2$ gauge nodes with color $\s$. }
\label{table:N=1}
\end{table}

Let us mention that the anomalies for the class $\CS$ theories having neither punctures nor twist lines can be obtained from integrating the anomaly polynomial for the 6d $\CN=(2, 0)$ theory \cite{Harvey:1998bx,Intriligator:2000eq,Yi:2001bz} as done in \cite{Alday:2009qq,Benini:2009mz, Bah:2012dg}.

\paragraph{Anomaly coefficients in the presence of the twist line}

Suppose we have a UV curve $\CC_{g, n}$ with the normal bundles, having degrees $(p, q)$. Let us assume that all the punctures are of maximal type. Now, consider a colored pair-of-pants decomposition such that the number of $\CN=2$ gluing with $+$ color is $x$, the number of $\CN=2$ gluing with $-$ color is $y$ and the number of $\CN=1$ gluing is $z$. This leads to the following anomaly coefficients: 
\begin{align}
\begin{split}
\tr J_{+}&= \tr J_{+}^{3} = q(r_{\G}-3d_{\G})+(2y+z)d_{\G}\ ,\\
\tr J_{-}&= \tr J_{-}^{3}= p(r_{\G}-3d_{\G})+(2x+z)d_{\G}\ ,\\
\tr J_{+}J_{-}^{2}&=2q\left(\frac{2}{3} h^\vee_\G d_\G +\frac{1}{2}r_\G - \frac{3}{2}d_\G\right)+(2x+z)d_{\G} \ , \\ 
\tr J_{+}^{2}J_{-}&=2p\left(\frac{2}{3} h^\vee_\G d_\G +\frac{1}{2}r_\G - \frac{3}{2}d_\G\right)+(2x+z)d_{\G} \ . 
\end{split}
\end{align}
Now we use following identities:\footnote{If a gluing is done between two pairs-of-pants labelled by $\sigma_{i}$ and $\sigma_{j}$, let us assign the value $F_{\langle ij\rangle}=\sigma_{i}+\sigma_{j}$ to the gluing, denoted by $\langle ij\rangle$. Now we will consider the sum $I=\sum_{ij}\left(\sigma_{i}+\sigma_{j}\right)$ in two different ways. Since each pair-of-pants has 3 punctures, $I=3(p-q)$. On the other hand $I=\sum_{\textrm{gluing}} F_{\langle ij\rangle}$, now $F_{\langle ij\rangle}$ is $0$ if we have $\mathcal{N}=1$ gluing, $F_{\langle ij \rangle}=\pm 2$ respectively for gluing two $\pm$ type pairs-of-pants. Hence, $I= 2x-2y+0$ and $I =2(x-y)$. So we have $2(x-y)=3(p-q)$. Note that if we do not weight the gluing by their signs, then we would have obtained $2(x+y+z)=3(p+q)$.}
\begin{align}
(2x+z)+(2y+z)=3(p+q),\ \qquad (2x+z)-(2y+z)=3(p-q).
\end{align}
to recast the anomaly coefficients as:
\begin{align}
\begin{split}
\tr J_{+}&=\tr J_{+}^{3} = qr_{\G}\ ,\\
\tr J_{-}&=\tr J_{-}^{3} = pr_{\G}\ ,\\
\tr J_{+}J_{-}^{2}&=2q\left(\frac{2}{3} h^\vee_\G d_\G +\frac{1}{2}r_\G\right) \ , \\
\tr J_{+}^{2}J_{-}&=2p\left(\frac{2}{3} h^\vee_\G d_\G +\frac{1}{2}r_\G\right)\ .
\end{split}
\end{align}

Now, let us introduce a twist loop passing through a cycle on the UV curve. Upon colored pair-of-pants decomposition, it will pass through number of $+$ type and $-$ type pants. Let us write the sequence of the color of the pants as $\s_1 \s_2 \cdots \s_{\ell}$. The sequence is cyclic so that $\s_{\ell+1} = \s_1$. As the twist line passes through the pairs-of-pants, it replaces $T_\G^{\s_i}$ by $\tilde{T}_\G^{\s_i}$ and also changes the gauge group $\G^{\s_i \s_{i+1}}$ between the pants $i$ and $j$ by $G^{\s_i \s_{i+1}}$. Here $+-$ and $-+$ refer to $\CN=1$ vector multiplet, whereas $++$ and $--$ refer to $\CN=2$ vector. Now, the effect of the twist line to the anomalies can be schematically written as
\begin{align}
\begin{split}
 \delta_{\CA} &= \sum_{i=1}^{\ell} \left(\tilde{T}_\G^{\s_i} - T_\G^{\s_i} + G^{\s_i \s_{i+1}} - \G^{\s_i \s_{i+1}} \right) \\
 &= \sum_{i=1}^{\ell} \left[ \half \left( \tilde{T}_\G^{\s_i} - T_\G^{\s_i} \right) + \half \left( \tilde{T}_\G^{\s_{i+1}} - T_\G^{\s_{i+1}} \right) + \left( G^{\s_i \s_{i+1}} - \G^{\s_i \s_{i+1}} \right) \right] \ , 
\end{split}
\end{align}
where each symbol means the anomaly coefficients of the corresponding block. It is easy to verify that the term in the square bracket vanishes. Therefore adding a twist loop does not alter the anomaly coefficients. 

\paragraph{Anomaly coefficients in the presence of punctures}

Let us derive the formula for the anomaly coefficients in the presence of $n_{\G\pm}/n_{G\pm}$ number of $\G$/$G$-type punctures (untwisted/twisted full punctures) with color $\pm$. We define $n_{\pm}=n_{G\pm}+n_{\G\pm}$. While $n_{G\pm}$ is an even number, $n_{\G\pm}$ can be even or odd. Our argument about the twist loop still holds in the presence of the puncture, so the result is not affected by the twist. 

When $n_{\G\pm}$ are all even, we can glue all the punctures pairwise with $\mathcal{N}=2$ vector multiplets in some particular dual frame. This yields the UV curve with no punctures with the degrees of the line bundles given by $(p+n_{+}, q+n_{-})$. Now we can calculate the anomalies corresponding to this modified surface and then subtract the contribution from the vector multiplets due to the gluing. In this way, we obtain
\begin{align}
\label{Anomaly}
\begin{split}
\tr J_+ &= \tr J_{+}^{3}=(q+n_{-})r_{\G}-2n_{\G-}d_{\G}-2n_{G-}d_G\ ,\\
\tr J_- &= \tr J_{-}^{3}=(p+n_{+})r_{\G}-2n_{\G+}d_{\G}-2n_{G+}d_G\ ,\\
\tr J_{+}^{2}J_{-}&=2(p+n_{+})\left(\frac{2}{3} h^\vee_\G d_\G +\frac{1}{2}r_\G\right)-2n_{\G+}d_{\G}-2n_{G+}d_G\ , \\
\tr J_{+}J_{-}^{2}&=2(q+n_{-})\left(\frac{2}{3} h^\vee_\G d_\G +\frac{1}{2}r_\G\right)-2n_{\G-}d_{\G}-2n_{G-}d_G.
\end{split}
\end{align}

\begin{figure}[!h]
\begin{center}
  \includegraphics[width=6in]{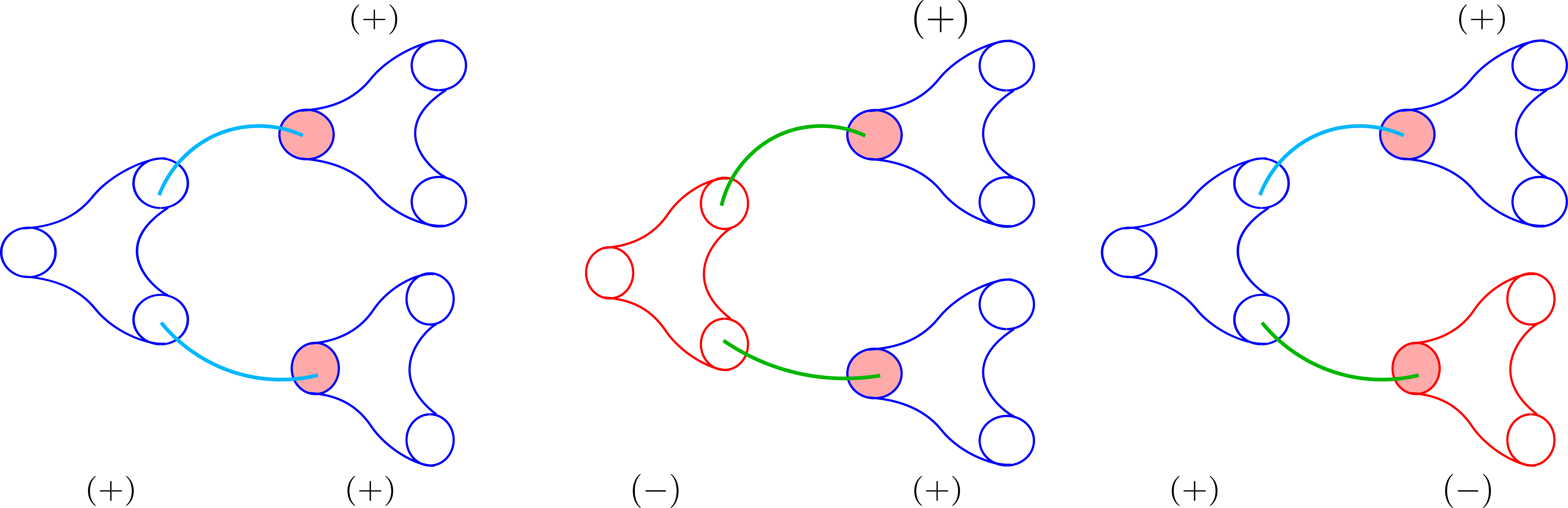}
   \end{center}
 \caption{Inclusion of an extra pair-of-pants (left most one in each figure) to the UV curve when $n_{\G\pm}$ is odd. The solid lines denote extra gluing. Green lines denote $\mathcal{N}=1$ gluing while the blue lines are for the $\mathcal{N}=2$ gluing.  
 There are 3 more diagrams we can have, which are obtained by swapping the colors of pairs-of-pants from blue to red and vice-versa in each of the diagrams.}
 \label{Fig:N1}
\end{figure}

If any or both of $n_{\G\pm}$ is odd, then we can consider $n_{\G\pm}^{\prime}\equiv n_{\G\pm}-1$ punctures, for whichever one ($+$ or $-$ or both) is odd. Since, $n_{\G\pm}^{\prime}$ is an even number, we can employ the previous trick to compute the anomaly coefficients, which we obtain by substituting $n_{\pm}\goto n_{\pm}^{\prime}\equiv n_{\pm}-1$. In order to get the correct anomaly coefficient, all we have to do is to add the extra pair-of-pants with appropriate label to the rest of the diagram. This is effectively the same as adding a pair-of-pants to the diagram with a $\mathcal{N}=2$ vector multiplet with both of them having the same color assignment as the puncture. Hence, the formula given above by equation \eqref{Anomaly} holds true for this case as well. 
The diagrams given in the figure \ref{Fig:N1} will elucidate that either we get explicitly an extra $\mathcal{N}=2$ gauge node of the same color as the puncture or we get two extra $\mathcal{N}=1$ vector multiplet while losing one $\mathcal{N}=2$ vector multiplet of opposite color of puncture. It can be easily verified that adding two extra $\mathcal{N}=1$ vector multiplets while losing one $\mathcal{N}=2$ vector multiplet of opposite sign of the puncture is effectively the same as adding one $\mathcal{N}=2$ vector multiplet of the same color as the puncture. All of these different gluing represent dual frames of the same theory.

The punctures also introduce flavor symmetries, hence we have more anomaly coefficients. For a maximal puncture carrying the flavor symmetry group $\mathfrak{g}$ and labelled by the color $\sigma$ we have 
\begin{align}
\Tr J_{\pm}T^{a}T^{b}=-\frac{1\mp\sigma}{2}h^{\vee}_{\mathfrak{g}}\delta^{ab} . 
\end{align}
Here $\mathfrak{g}=\G$ if the puncture is untwisted and $\mathfrak{g}=G$ if the puncture is twisted.

%

\subsection{$a$-maximization}
The superconformal $R$ symmetry in infrared is given by:
\begin{align}
R_{IR}=R_{0}+\epsilon\mathcal{F}
\end{align}
where $\epsilon$, hence $R_{IR}$ is determined by $a$-maximization \cite{Intriligator:2003jj}. 
The $a$-function is given by the trace anomalies as 
\begin{align}
a=\frac{3}{32}\left(3\tr R_{IR}^{3}-\tr R_{IR}\right) \ . 
\end{align}
We determine the value of $\e$ by using the trial $R$-charge above and then solving
\begin{align}
 \frac{\p a}{\p \e} \bigg|_\e = 0 \ , \quad \frac{\p^2 a}{\p \e^2} \bigg|_\e  < 0 \ . 
\end{align}
If two theories have identical anomaly coefficients, it would imply that the polynomial $a(\epsilon)$ is the same, subsequently, we have the same $\epsilon$ as a maximizing solution. The only thing that we are required to check is whether there is any accidental symmetry appearing from the decoupling of any operators since the operator content of the two theories of interest might be different. Should we have any operator violating the unitarity bound, that gets decoupled, and we are required to do the $a$-maximization once again upon removing the contributions coming from those operators violating the unitary bound \cite{Kutasov:2003iy}. 

In our case, as long as $p>0,q>0$, it is sufficient to verify whether $|\epsilon|\leq\frac{1}{3}$, since the chiral operators of lowest $R$-charge have $(J_+, J_-)=(2, 0)$ or $(0, 2)$. 
\begin{figure}[!h]
\centering
\includegraphics[width=2.5in]{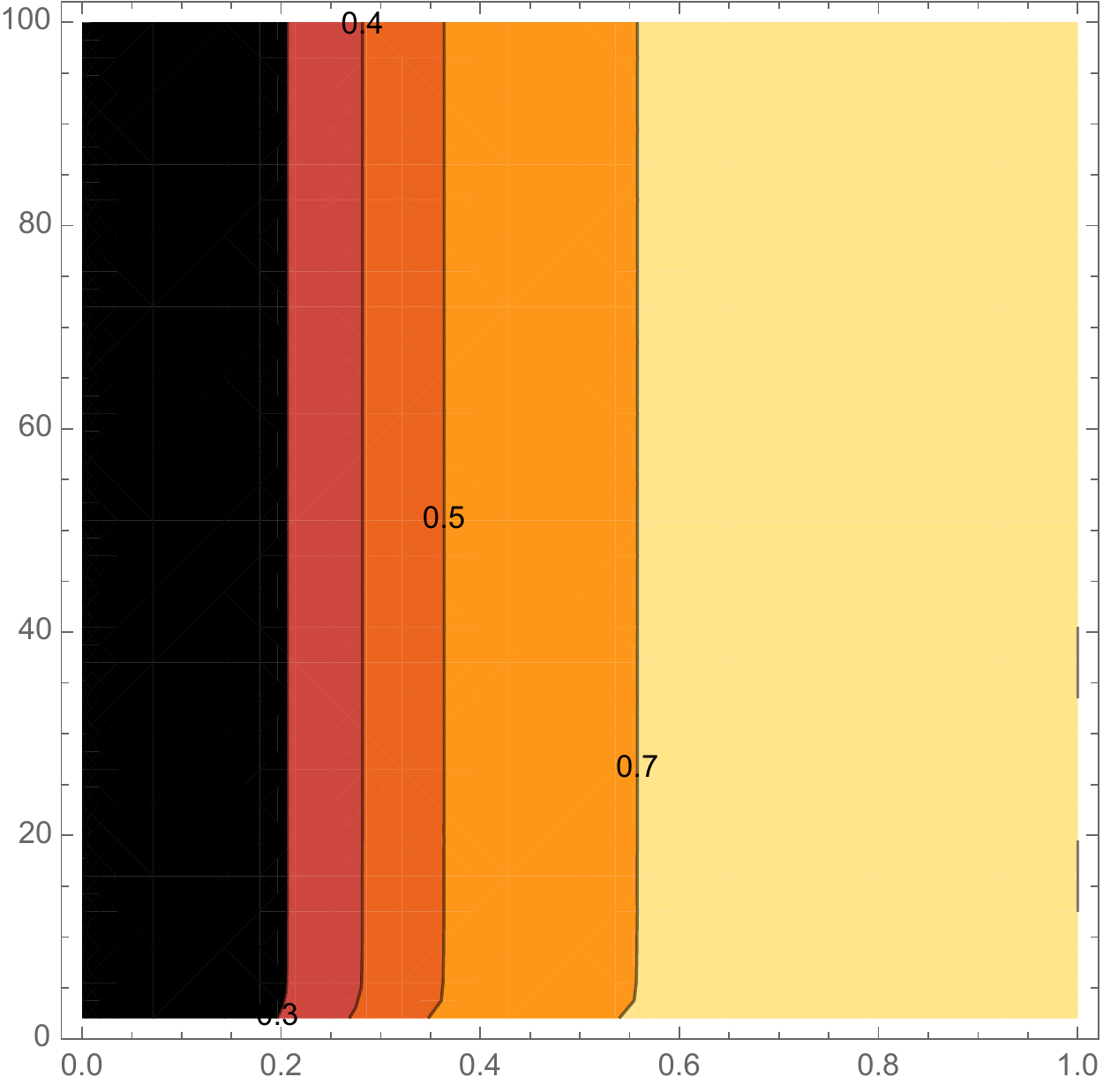} \quad
\includegraphics[width=2.5in]{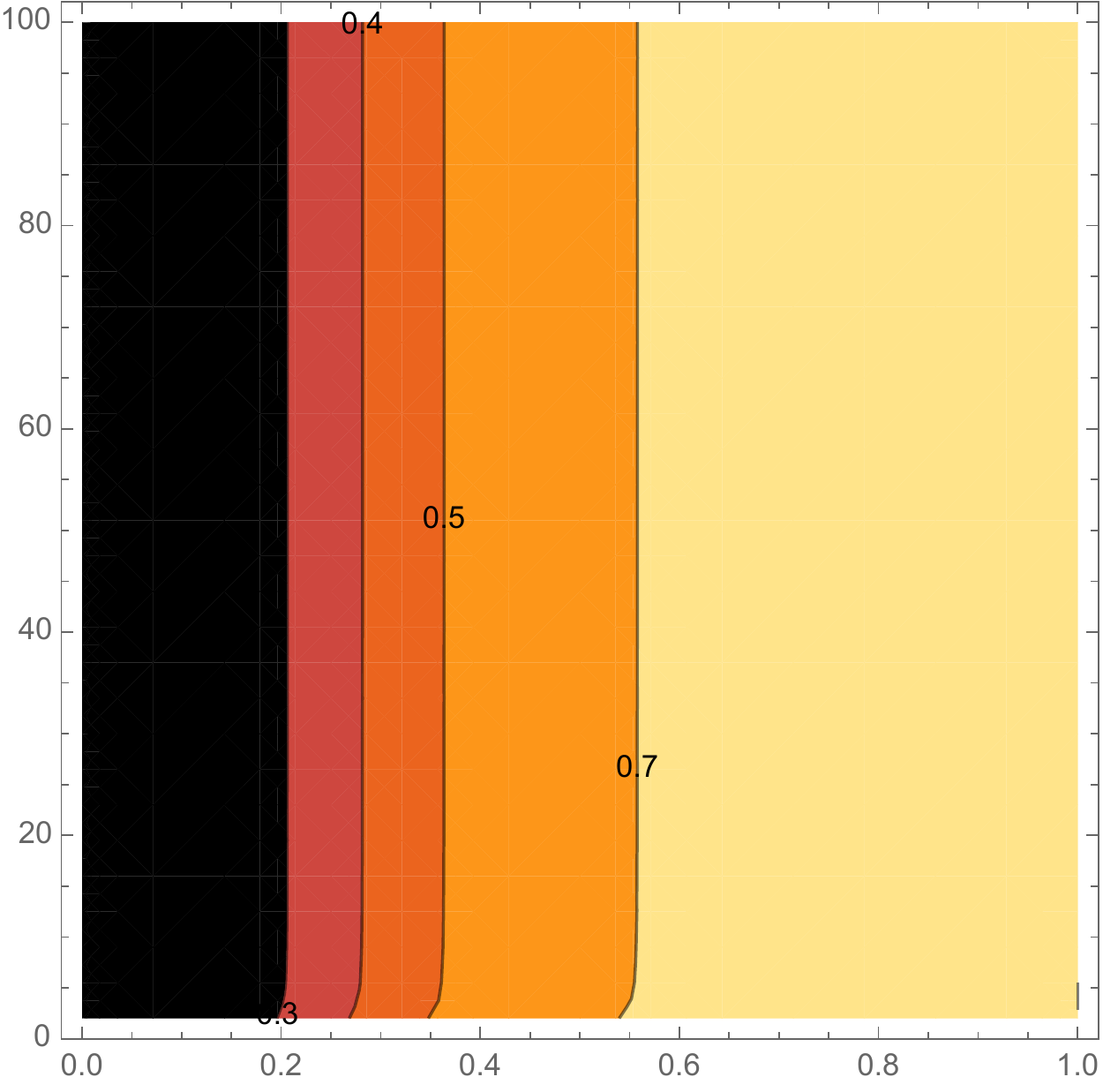}
\caption{A contour plot of $3\epsilon$ that maximizes $a(\epsilon)$ for $\G=D_{n}$ and $\G=A_{n-1}$. The $Y$ axis denotes $n$, and the $X$ axis is for the $t=\frac{p-q}{p+q}$, parametrizing the choice of the normal bundles. Note the $t$ has to be in this range $|t|\leq1$. Since, $\epsilon(t)=-\epsilon(t)$, it is sufficient to plot for $0\leq t\leq 1$.}
\label{fig:amax}
\end{figure}
We find that for a UV curve $\CC_{g,0}$ with degrees of line bundle, given by $p$ and $q$, the trial $a$-function comes out to be: 
\begin{equation}
a(\epsilon)=-\frac{9}{32} d_\G h^\vee_\G x \epsilon^2+\frac{9 d_{\G}h^{\vee}_{\G}x}{32}-\frac{9d_{\G}h^\vee_\G y}{32}\epsilon ^3+\frac{9d_{\G}h^\vee_\G y}{32} \epsilon +\frac{3 r_{\G} x}{16}-\frac{9r_{\G}y}{32}\epsilon^3+\frac{3r_{\G}y\epsilon}{32}
\end{equation}
where $x=\frac{p+q}{2}$ and $y=\frac{p-q}{2}$. This is true irrespective of presence/absence of twist line loops on the UV curve. Here $d_{\G}$ is the dimension of flavor symmetry group $\G$ associated with untwisted puncture and $h^{\vee}_{\G}$ is the dual Coxeter number of $\G$.
We define $g_\G=\frac{r_\G}{d_\G h^\vee_\G}$ where $r_{\G}$ is the rank of the group $\G$ and $t=\frac{p-q}{p+q}$. The $\epsilon$ that maximizes the trial $a$-function is given by
\begin{equation}
\epsilon(t)=\frac{\sqrt{(g_\G+1) (g_\G+3) t^2+1}-1}{3 (g_\G+1) t}
\end{equation}
For $A_{n-1}$, we have $\frac{1}{g_\G}=n(n+1)$, for $D_n$ we have $\frac{1}{g_\G}=(2n-1)(2n-2)$; for $E_{6}$, we have $\frac{1}{g_\G}=156$.
For each of these case, we find $|\epsilon|\leq\frac{1}{3}$ as long as $p>0, q>0$. See the figure \ref{fig:amax} for the case of $\G=D_{n}$ and $A_{n-1}$. Therefore none of the operators decouple along the renormalization group flow as long as $p>0, q>0$ with no punctures. 

\subsection{Superconformal index} \label{sec:N2idx}
In this section, we discuss the superconformal indices of the class $\CS$ theories that we studied so far. The superconformal index is a supersymmetric partition function that captures the multiplicities of the short multiplets in a SCFT. It provides a strong check of the supersymmetric dualities. We remark that most of discussion in the current subsection is already known. We would simply like to highlight the effect of the twist lines in the computation of the index. We refer to the original papers \cite{Gadde:2009kb,Gadde:2011ik,Gadde:2011uv,Gaiotto:2012xa, Beem:2012yn} and the review article \cite{Rastelli:2014jja} and the references therein for more detail. 
We would like to focus on the effect of the outer-automorphism twist in the index, which has been discussed in \cite{Lemos:2012ph, Mekareeya:2012tn, Agarwal:2013uga, Chacaltana:2015bna}.

The superconformal index for an $\CN=2$ class $\CS$ theory can be written as a partition function of certain 2d topological field theory (TQFT) on the UV curve $\CC_{g, n}$ of genus $g$ and $n$ punctures (here we assume all the punctures to be maximal) as
\begin{align} \label{eq:N2idx}
 \CI_{\CC_{g, n}}(p, q, t; \vec{a}) = \sum_{\vec \lambda \in R_\G} C_{\vec \lambda}(p, q, t)^{2g-2+n} \prod_{I=1}^n \psi^{(I)}_{\vec \lambda}(p, q, t; \vec{a}_I)\ , 
\end{align}
where $C_{\vec\lambda}$ is called the structure constant of the TQFT and $\psi_{\vec\lambda}^{(I)}$ are the wave functions associated to the punctures. The parameters $p, q, t$ are the fugacities associated to some linear combinations of the Cartans of the superconformal group $SU(2, 2|2)$, and $\vec{a}_I$ are the fugacities associated to the global symmetry coming from each punctures. Notice that the sum is over all the representations $R_\G$ of the group $\G$ labeling the class $\CS$ theories. So far, we have assumed that there is no twist lines on the UV curve. 

For $g=0$, $n=3$, we obtain the index for the $T_\G$ theory given as
\begin{align}
 \CI_{T_\G}(\vec{a}_1, \vec{a}_2, \vec{a}_3) = \sum_{\vec\lambda \in R_\G} C_{\vec \lambda} \psi_{\vec\lambda}(\vec{a}_1) \psi_{\vec\lambda}(\vec{a}_2 ) \psi_{\vec\lambda}(\vec{a}_3) \ , 
\end{align}
where we omitted the dependence on the fugacities $p, q, t$. The wave functions satisfy 
\begin{align}
 \oint [d\vec{z}]_\G I_{\textrm{vec}, \G} (\vec{z}) \psi_{\vec\lambda} (\vec{z}) \psi_{\vec\mu}(\vec{z}) = \delta_{\vec \lambda \vec \mu} \ , 
\end{align}
where $I_{\textrm{vec}, \G}$ is the index for the vector multiplets and $[d\vec{z}]_\G$ is the Haar measure for the gauge group $\G$. From this orthonormality condition, one can easily obtain the index formula \eqref{eq:N2idx} for the arbitrary UV curve $\CC_{g,n}$ by multiplying the contributions from each block and the vector multiplets and then by integrating over the gauge groups. The formula \eqref{eq:N2idx} is independent of the pair-of-pants decomposition, which is consistent with the duality. 

When there is a twist line, we have $\tilde{T}_\G$ blocks along with the $T_\G$ theories. The index for the $\tilde{T}_\G$ is given by
\begin{align}
 \CI_{T_G}(\vec{a}, \vec{b}_1, \vec{b}_2) = \sum_{\vec\lambda \in R_G} C_{\vec \lambda} \psi_{\vec\lambda}(\vec{a}) \tilde{\psi}_{\vec\lambda}(\vec{b}_1 ) \tilde{\psi}_{\vec\lambda}(\vec{b}_2) \ , 
\end{align}
where the wave functions are changed appropriately to that of the twisted punctures. Here the summation is over the representations $R_G$ of the group $G$, instead of $\G$. This means that for the untwisted punctures $\psi_{\vec\lambda}$, we sum over only over the representations of $\G$ that are invariant under the $\IZ_2$ outer-automorphism action. The wave functions for the twisted punctures are orthonormal under the vector multiplet measure for the gauge group $G$:
\begin{align}
 \oint [d\vec{z}]_G I_{\textrm{vec}, G} (\vec{z}) \tilde{\psi}_{\vec\lambda} (\vec{z}) \tilde{\psi}_{\vec\mu}(\vec{z}) = \delta_{\vec \lambda \vec \mu }
\end{align}
Now, one can easily obtain the index in the presence of a twist line as
\begin{align} \label{eq:N2idxtw}
 \CI_{\tilde{\CC}_{g, n}} (p, q, t; \vec{a}) = \sum_{\vec \lambda \in R_G} C_{\vec \lambda}(p, q, t)^{2g-2+n} \prod_{I=1}^n \psi^{(I)}_{\vec \lambda}(p, q, t; \vec{a}_I)\ , 
\end{align}
when all the external punctures are untwisted. Notice that the only difference between \eqref{eq:N2idx} and \eqref{eq:N2idxtw} is the domain of summation. This clearly implies that the indices for the theory with the twist line differs from the one without the twist line. Also, there is no particular choice of the twist loop either. This is consistent our claim that there is only one physically inequivalent configuration of the twist loop. 

Let us consider some simple examples to see this explicitly. Consider the Schur limit of the index, which takes $p \to 0, q=t$. Then the wave function and the structure constant can be written as 
\begin{align} 
 \psi_{\vec\lambda} (\vec{a}) = \PE \left[\frac{q}{1-q} \chi_{\textrm{adj}} (\vec{a}) \right] \chi_{\vec\lambda}(a) \ , \qquad C_{\vec\lambda} = \prod_{i=1}^{r_\G} (q^{d_i}; q) / \chi_{\vec\lambda}(q^{\vec\rho}) \ , 
\end{align}
where PE denotes the plethystic exponential, and the $q$-Pochammer symbol is given by $(x; q)\equiv \prod_{n\ge0} (1-xq^n)$, and $d_i$ are the degrees of the Casimirs of $\G$ and $r_\G$ is the rank of $\G$. $\chi_{\vec\lambda}(\vec{z})$ is the character of the representation $\vec\lambda$ and $\vec\rho$ denotes the Weyl vector of $\G$ and $q^{\vec\rho} \equiv (q^{\rho_1}, q^{\rho_2}, \ldots, q^{\rho_{r_\G}})$. 

For example, when $g=2, n=0$ and $\G = D_4$, we obtain
\begin{align}
\CI_{\CC_{2, 0}} (q) = 1-2 q^2-2 q^3-5 q^4-2 q^5+8 q^6+8 q^7+24 q^8+O(q^9) \ , 
\end{align}
for the case without the twist loop. When we have a twist loop, the sum is over the representations of $G=C_3$. We get
\begin{align}
 \CI_{\tilde{\CC}_{2, 0}}(q) = 1-2 q^2-2 q^3-5 q^4-2 q^5+6 q^6+12 q^7+26 q^8+O(q^9) \ . 
\end{align}
The indices for the two theories indeed differ, and they start to deviate at order 6. 

Let us consider another example: $g=1, n=1$ and $\G=D_4$. For this case, we get
\begin{align}
\begin{split}
 \CI_{\CC_{1, 1}} (q; \vec{a}) &= 1 + q \chi_{[0100]} + q^2 \left(\chi_{[0100]} + \chi_{[2000]} + \chi_{[0020]} + \chi_{[0002]} + \chi_{[0200]} \right) \\
 &~~+ \big(\chi_{[0300]}+\chi_{[2100]}+\chi_{[0102]}+\chi_{[0120]}+\chi_{[0200]}+2\chi_{[1011]}+\chi_{[0020]} \\
 & \qquad+\chi_{[2000]}+\chi_{[0002]}+4 \chi_{[0100]} +\chi_{[1000]}+ \chi_{[0010]}+ \chi_{[0001]} \big)q^3 + O(q^4) \\
 &= 1+28 q+433 q^2+4867 q^3 + 44234 q^4+343744 q^5+O(q^6) \ , 
\end{split}
\end{align}
for the case without the twist loop. Here $\chi_R$ denotes the character of the $SO(8)$ for the representation given by the Dynkin label $R$. The final line is the result after taking the flavor fugacities $\vec{a}$ to 1. On the other hand, when the UV curve has a twist loop, we get
\begin{align}
\begin{split}
 \CI_{\tilde{\CC}_{1, 1}}(q; \vec{a}) &=  1 + q \chi_{[0100]} + q^2 \left(\chi_{[0100]} + \chi_{[2000]} + \chi_{[0020]} + \chi_{[0002]} + \chi_{[0200]} \right) \\
 &~~+ \big(\chi_{[0300]}+\chi_{[2100]}+\chi_{[0102]}+\chi_{[0120]}+\chi_{[0200]}+2\chi_{[1011]}+\chi_{[0020]} \\
 & \qquad+\chi_{[2000]}+\chi_{[0002]}+4 \chi_{[0100]} +\chi_{[1000]} \big)q^3 + O(q^4) \\
 &= 1+28 q+433 q^2+4851 q^3+43802 q^4+337264 q^5+O(q^6) \ .
\end{split}
\end{align}
The difference begins at order $q^3$. The linear term in the index comes from the conserved current multiplet. Therefore, the global symmetries for both theories are identical and there is no symmetry enhancement from $SO(8)$ which comes from the puncture.

Let us comment that our discussion can be easily generalized to $\CN=1$ class $\CS$ theories. Especially, the $\CN=1$ index also admits the Schur limit, which gives the identical partition function as the $\CN=2$ counterpart. In this limit, the dependence on the degrees of the normal bundles vanishes. 

Therefore we see that the theory given by the UV curve $\CC_{g, n}$ and the identical curve with the twist loop $\tilde{\CC}_{g, n}$ have different superconformal indices, but the same anomalies. 
Also, we notice that whenever the twisted puncture is introduced, it changes the range of summation from the representations of $\G$ to $G$. Therefore the index does not altered further upon inserting a new loop of twist line. This is consistent with the duality as stated in \eqref{dual}. 


\section{Conclusion}

In this paper, we have studied four-dimensional superconformal theories in class $\CS$, in the presence of the outer-automorphism twist line forming a loop on the UV curve. We have found new dualities in the presence of the twist loop and performed checks using anomalies and superconformal index.  Quite interestingly, we have found that the theory with a loop of twist line on the UV curve and the theory without the twist line have the identical anomalies, but the indices differ. This provides a simple way of constructing a pair of distinct theories giving identical 't Hooft anomalies. 

The examples of misleading anomaly matchings we find are somewhat analogous to the orbifolds in 2d CFTs. As is well-known, orbifolding does not change the central charges of the theory but it does change the theory in a non-trivial way. Our example can be thought of as a higher-dimensional analogue of orbifolding. When we obtain the 4d SCFT, we truncate certain degrees of freedom that are not invariant under the outer-automorphism twist. We also get some extra states similar to the twisted sector. But this operation does not change the overall behavior regarding the growth of the number of states. 
We expect that it is straight-forward to generalize our observations to the theories having some higher dimensional origin. For example, one can consider 4d $\CN=1$ or 2d $\CN=(0, 2)$ theories coming from 6d $\CN=(1, 0)$ theories on a Riemann surface \cite{Gaiotto:2015usa, Franco:2015jna,Hanany:2015pfa,DelZotto:2015rca,Heckman:2016xdl, Razamat:2016, Putrov:2015jpa} with a twist loop.  

In this paper, we have found pairs of theories having the identical anomalies. One interesting question to ask is whether it is possible to have many more theories with identical anomalies. In our examples, the theory has exactly marginal deformations which does not change the anomalies, so technically speaking there are actually infinitely many theories giving the same anomalies. But once we mod out by such exactly marginal deformations, how many distinct theories are there for given anomalies? We were able to distinguish them using the superconformal indices. Can there be another theory giving a different index function with the same anomalies?  The lens space index provides even more refined quantity, which can distinguish the theories having different set of non-local operators \cite{Aharony:2013hda, Razamat:2013opa}. How many theories (up to exactly marginal deformations) with distinct lens space indices exist for a given anomalies? It would be interesting to answer these questions.  

Finally, we would like to point out that all the examples of misleading anomaly matching we studied are `non-Lagrangian' with no known Lagrangian description. It would be interesting to find more examples of conventional Lagrangian gauge theory or find if there is any obstruction to such cases. 

\acknowledgments
We would like to thank Ken Intriligator, Emily Nardoni and Yuji Tachikawa for discussions and correspondence. This work is supported by the US Department of Energy under UCSD's contract de-sc0009919. The work of JS is also supported in part by Hwa-Ahm foundation.

\bibliographystyle{jhep}
\bibliography{refs}
\end{document}